\documentclass[letterpaper,aps,prd,onecolumn,superscriptaddress,showpacs,preprintnumbers,nofootinbib]{revtex4}
\usepackage{graphicx}
\usepackage[intlimits,centertags]{amsmath}
\usepackage{amssymb,amsfonts}
\usepackage{gensymb}
\usepackage{mathrsfs}
\usepackage{hyperref}

\begin{document}

\title{Analytic Solutions of Ultra-High Energy Cosmic Ray Nuclei Revisited}

\author{Markus~Ahlers} 
\affiliation{C.N.Yang Institute for Theoretical Physics, SUNY at Stony Brook, Stony Brook, NY 11794-3840, USA}

\author{Andrew~M.~Taylor}
\affiliation{ISDC, Chemin d'Ecogia 16, Versoix, CH-1290, Switzerland}

\begin{abstract}
The chemical composition of ultra-high energy cosmic rays is a key question in particle astrophysics. The measured composition, inferred from the elongation rates of cosmic ray showers, looks in general very different from the initial source composition: resonant photo-disintegration in the cosmic radiation background proceeds rapidly at the highest energies and the initial composition quickly becomes lighter during propagation. For a statistical analysis of continuously improving cosmic ray data it is desirable to know the secondary spectra as precisely as possible. Here, we discuss exact analytic solutions of the evolution equation of ultra-high energy cosmic ray nuclei. We introduce a diagrammatic formalism that leads to a systematic analytic expansion of the exact solution in terms of second order effects of the propagation. We show how the first order corrections of this expansion can improve the predictions of secondary spectra in a semi-analytical treatment.
\end{abstract}


\preprint{YITP-SB-10-33}

\maketitle

\section{Introduction}

The mass composition of ultra-high energy (UHE) cosmic rays (CRs) remains an open question in astrophysics. The average mass number $\langle A\rangle$ per energy can be inferred {\it directly} in CR observatories by the measurement of the elongation rate distribution of CR showers~\cite{Nagano:2000ve}. Presently, the observational situation is ambiguous despite strong experimental efforts over the years. Recent findings of the Pierre Auger collaboration~\cite{Abraham:2010mj} indicate a transition of UHE CRs within the energy range $10^{18}$~eV to $4\times10^{19}$~eV from a light (presumably proton-dominated) spectrum towards a heavier composition~\cite{Abraham:2010yv}. In contrast, the HiRes collaboration~\cite{Abbasi:2007sv} finds a mass composition compatible with that of a proton-dominated spectrum~\cite{Abbasi:2009nf}.

Various features in the CR spectrum can also provide {\it indirect} evidence for the origin and composition of UHE CRs. The {\it ankle} -- a hardening of the spectrum at $3\times10^{18}$~eV -- could be formed naturally by the superposition of two power-law fluxes and serves as a candidate of the transition between galactic and extra-galactic cosmic rays~\cite{Linsley:1963bk,Hill:1983mk}. It has also been advocated that this feature could be well reproduced by a proton-dominated power-law spectrum, where the ankle is formed as a {\it dip} in the spectrum from the energy loss of protons via Bethe-Heitler pair production~\cite{Berezinsky:2002nc,Fodor:2003ph}. In this case extra-galactic protons could already start to dominate the spectrum beyond the {\it 2nd knee} which corresponds to a slight {\it softening} of the spectrum at $5\times10^{17}$~eV. 

Proton-dominance beyond the ankle is ultimately limited by the {\it Greisen-Zatspin-Kuzmin} (GZK) cutoff~\cite{Greisen:1966jv,Zatsepin:1966jv} due to resonant photo-pion production in the cosmic microwave background (CMB). In fact, a suppression of the CR spectrum at the expected energy of about $5\times10^{19}$~eV has been detected by the Pierre Auger and HiRes collaborations at a statistically significant level~\cite{Abbasi:2007sv,Abraham:2008ru} and is consistent with a proton dominance at these energies. However, this feature could also originate from photo-disintegration of UHE CR nuclei  in the cosmic background radiation, or from an {\it in situ} energy cut-off of the injection spectrum of UHE CR. To summarize, the interpretation of these experimental findings is as yet inconclusive and even controversial. 

Simple theoretical arguments, however, can motivate a significant contribution of primary nuclei at energies beyond $10^{18}$~eV. For the efficient acceleration of primary CRs to these extreme energies a particle should be confined magnetically in a suitable astrophysical environment. Since the particle's Larmor radius is proportional to its rigidity, {\it i.e.}~its energy per charge, we expect that the maximal energy $E_{\rm max}$ of UHE CRs to scale with the charge number $Z$ of a (fully ionized) nucleus. The acceleration of heavy nuclei like iron ($Z=26$) can hence proceed up to larger energies and alleviates the fundamental limitations of cosmic accelerators to account for the observed spectrum of UHE CRs~\cite{Amsler:2008zzb}.

Analytic\footnote{We use the term ``analytic'' here to denote {\it explicit} analytic solutions in closed form following~\cite{Hooper:2008pm}. There also exist many {\it implicit} analytic solutions for the spectra of UHE CRs that require an algorithmic treatment, {\it e.g.}~\cite{Berezinsky:2002nc,Aloisio:2008pp,Ahlers:2009rf}.} descriptions of UHE CR propagation provide an easily accessible means of exploring both proton and heavy nuclei source scenarios. With the most recent results of the Auger collaboration indicating that the composition continues to become heavier at energies above $3\times 10^{19}$~eV, the heavy UHE CR flux component may well arrive from very local cosmological regions. To facilitate the future exploration of nearby UHE CR nuclei source distributions, we here develop further the analytic description of UHE CR nuclei propagation put forward in Ref.~\cite{Hooper:2008pm}. These developments take into account subdominant energy losses, ensuring that this description provides an accurate means of obtaining the UHE CR flux over the full energy range covered by the cosmic ray observatories. 

We will start in section~\ref{sec:propagation} with a short recapitulation of the compact evolution equations of UHE CR nuclei in the limit of a spatially homogeneous distribution of isotropic CR sources. We derive an exact analytic solution in section~\ref{sec:analytic} that includes continuous energy losses and multi-nucleon loss transitions between nuclei. In section~\ref{sec:perturbation} we introduce a perturbative expansions of the exact analytic solution that provides a convenient practical framework for next-to-leading order corrections of the solution given in Ref.~\cite{Hooper:2008pm}. We finally conclude in section~\ref{sec:conclusions}.

\section{Propagation of Cosmic Ray Nuclei}\label{sec:propagation}

For a spatially homogeneous distribution of cosmic sources, emitting UHE particles of type $i$, the co-moving number density $Y_i$ is governed by a set of (Boltzmann) continuity equations of the form:
\begin{equation}\label{eq:diff0}
\dot Y_i = \partial_E(HEY_i) + \partial_E(b_iY_i)-\Gamma^{\rm tot}_{i}\,Y_i
+\sum_j\int{\rm d} E_j\,\gamma_{ji}Y_j+\mathcal{L}_i\,,
\end{equation}
together with the Friedman-Lema\^{\i}tre equations describing the cosmic expansion rate $H(z)$ as a function of red-shift
$z$.\footnote{This is given by \mbox{$H^2 (z) = H^2_0\,[\Omega_{\rm m}(1 + z)^3 + \Omega_{\Lambda}]$}, normalized to its value today of $H_0 \sim70$ km\,s$^{-1}$\,Mpc$^{-1}$, in the usual ``concordance model'' dominated by a cosmological constant with $\Omega_{\Lambda} \sim 0.7$ and a (cold) matter component, $\Omega_{\rm m} \sim 0.3$~\cite{Amsler:2008zzb}. The time-dependence of the red-shift can be expressed via ${\rm d}z = -{\rm d} t\,(1+z)H$.}  The first and second terms on the r.h.s.~describe, respectively, red-shift and other continuous energy losses (CEL) with rate $b \equiv \mathrm{d}E/\mathrm{d}t$.  The third and fourth terms describe more general interactions involving particle losses ($i \to$ anything) with total interaction rate $\Gamma^{\rm tot}_i$, and particle generation of the form $j\to i$ with differential interaction rate $\gamma_{ij}$.  The last term on the r.h.s., $\mathcal{L}_i$, corresponds to the emission rate of CRs of type $i$ per co-moving volume. 

The two main reactions of UHE CR nuclei during their cosmic evolution are photo-disintegration~\cite{Stecker:1969fw,Puget:1976nz,Stecker:1998ib,Goriely:2008zu} and Bethe-Heitler pair production~\cite{Blumenthal:1970nn} with the cosmic background radiation (CBR). The former process is dominated by the giant dipole resonance (GDR) with main branches $A\to(A-1)+N$ and $A\to(A-2)+2N$ where $N$ indicates a proton or neutron~\cite{Stecker:1969fw,Puget:1976nz,Stecker:1998ib}. The GDR peak in the rest frame of the nucleus lies at at about $20$~MeV for one-nucleon emission, corresponding to $E^A_{\rm GDR} \simeq A\times 2\times\epsilon^{-1}_{\rm meV}\times10^{10}$~GeV in the cosmic frame with photon energies $\epsilon = \epsilon_{\rm meV}$~meV. At energies below 10~MeV there exist typically a number of discrete excitation levels that can become significant for low mass nuclei. Above 30~MeV, where the photon wavelength becomes comparable or smaller than the size of the nucleus, the photon interacts via substructures of the nucleus. Out of these the interaction with quasi-deuterons is typically most dominant and forms a plateau of the cross section up to the photo-pion production threshold at $\sim145$~MeV. Bethe-Heitler pair production can be treated as a continuous energy loss process with rate $b_A(z,E) = Z^2b_p(z,E/A)$, where $b_p$ is the energy loss rate of protons~\cite{Blumenthal:1970nn}. The (differential) photo-disintegration rate $\Gamma_{A\to B}(E)$ ($\gamma_{A\to B}(E,E')$) is discussed in more detail in Appendix~\ref{sec:appI}. 

The evolution of the spectra proceeds very rapidly on cosmic time scales and the flux of secondary nuclei, $J$, looks generally quite different from the initial injection spectrum, $J_{\rm inj}$. The reaction network of nuclei depend in general on a large number of stable or long-lived isotopes. If the life-time of an isotope is much shorter than its photo-disintegration rate it can be effectively replaced by its long-lived decay products in the network~(\ref{eq:diff0}). Typically, neutron-rich isotopes $\beta$-decay to a stable or long-lived nucleus with the same mass number. In most cases there is only one stable nucleus per mass number below ${}^{56}$Fe with the exception of the pairs ${}^{54}$Cr/${}^{54}$Fe, ${}^{46}$Ca/${}^{46}$Ti, ${}^{40}$Ar/${}^{40}$Ca and ${}^{36}$S/${}^{36}$Ar (see Fig.~\ref{fig:PSB}). We follow here the approach of {\it Puget, Stecker and Bredekamp} (PSB)~\cite{Puget:1976nz} and consider only a single nucleus per mass number $A$ in the decay chain of primary iron ${}^{56}$Fe. This PSB-chain of nuclei linked by one-nucleon losses is indicated as a red arrow in Fig.~\ref{fig:PSB}.

As described earlier, CR nuclei that undergo rapid photo-disintegration with CMB photons carry a Lorentz factor of about $\gamma=2\times10^{10}$. We can only strictly neglect long-lived secondary isotopes from the reaction network if the nucleus lifetime in the cosmic frame, $\gamma\tau$, is much smaller than the inverse photo-disintegration rate, which is of the order of $(4/A)$~Mpc. This corresponds to nucleon life-times of less than a few minutes. Figure~\ref{fig:PSB} shows also isotopes below ${}^{56}$Fe with life-time larger than about one minute in addition to the nuclei of the PSB-chain. In general, there is a large number of isotopes that are sufficiently long-lived in the cosmic frame to take part in the photo-disintegration process. Fortunately, a large degeneracy of intermediate isotopes with equal mass number affects only very heavy nuclei. The photo-disintegration of these degenerate nuclei, dominated by collective excitations of nucleons like the GDR, mostly depend on the mass number $A$. The fluxes calculated for nuclei in the PSB-chain are expected to give a good representation of the total flux per mass number. Note that most of the analytic formulae that we are going to introduce in the following can be easily generalized to the case of the full reaction network including all isotopes.

Note, that the Boltzmann equations~(\ref{eq:diff0}) do not take into account the deflection of charged CR nuclei during their propagation through inter-galactic and galactic magnetic fields. The strength of inter-galactic magnetic fields is limited to the range $10^{-16}$G - $10^{-9}$G~\cite{Kronberg:1993vk,Neronov:1900zz} and suggested to be of ${\cal O}(10^{-12})$G by simulations of large-scale structure formation~\cite{Dolag:2004kp}. In fact, if synchrotron radiation during propagation is negligible and the source distribution is homogenous, Eq.~(\ref{eq:diff0}) provides a good approximation of the spectral evolution {\it even} for CRs having small rigidity which suffer large deflections~\cite{Aloisio:2004jda}. However, magnetic inhomogeneities on small scales will suppress the spectrum of CRs with Larmor radius $\ell_\mathrm{L}<\ell_\mathrm{d}$ where $\ell_{\rm d}$ is the characteristic distance between sources. It has been shown that for particularly strong inter-galactic magnetic fields of strength $\sim 1$~nG and coherence length of $\sim 1$ Mpc, the diffusive propagation of CR protons will start to affect the spectrum below about $10^9$~GeV if $\ell_\mathrm{d} \sim 50$~Mpc~\cite{Aloisio:2004fz}. Depending on the diffusion regime, this can suppress the proton flux at $10^8$~GeV by a factor of 3 to 100. Due to the dependence $\ell_\mathrm{L} \propto 1/Z$ we expect that for heavy nuclei diffusive propagation can in principle remain important up to the ankle. The results of this paper are based on solutions of Eqs.~(\ref{eq:diff0}) and assume that the contribution of inter-galactic or galactic magnetic fields can be neglected for the calculation of the UHE CR spectrum.

\section{Analytic Solution}\label{sec:analytic}

The secondary nuclei produced via photo-disintegration carry approximately the same Lorentz factor as the initial nucleus and the differential interaction rate in Eqs.~(\ref{eq:diff0}) can be approximated as $\gamma_{A\to B}(E,E') \simeq \Gamma_{A\to B}(E)\delta(E' - (B/A)E)$. It is hence convenient to express the energy of a nucleus with mass number $A$ and red-shift $z$ as $A(1+z)E$ where $E$ denotes the energy {\it per nucleon}. Introducing the CR density per co-moving volume and nucleon energy, $N_{A,i} \equiv \Delta E_i(1+z)AY_{A}(z,(1+z)AE_i)$, and corresponding emission rates, $Q_{A,i} \equiv A(1+z)\Delta E_i \mathcal{L}(z,A(1+z)E_i)$ we can re-write Eqs.~(\ref{eq:diff0}) in the compact form\footnote{This form of the differential equation holds for nuclei heavier than beryllium. We can easily compensate for the process ${}^9$Be $\to$ ${}^4$He + ${}^4$He + n of the PSB chain (see Appendix~\ref{sec:appI}) by re-defining $N_{A,i}' = N_{A,i}/2$ for $A=2,3,4$ and $N_{A,i}'=N_{A,i}$ for other nuclei.}
\begin{equation}\label{eq:diff1}
\dot N_{A,i} \simeq \Gamma^{\rm CEL}_{A,i+1}N_{A,i+1}-\Gamma^{\rm CEL}_{A,i}N_{A,i} - \sum_{B<A}\Gamma_{(A,i)\to(B,i)} N_{A,i}+ \sum_{B>A}\Gamma_{(B,i)\to(A,i)} N_{B,i} + Q_{A,i}\,,
\end{equation}
where we define the rates:
\begin{align}\label{eq:rates}
\Gamma^{\rm CEL}_{A,i} &=\Gamma_{(A,i)\to(A,i-1)} \equiv \frac{b_A(z,A(1+z)E_i)}{A(1+z)\Delta E_i}\,,
&\Gamma_{(A,i)\to (B,i)} &\equiv \Gamma_{A\to B}(z,A(1+z)E_i)\,.
\end{align}

Hooper {\it et al.}~\cite{Hooper:2008pm} discussed an analytical solution of Eqs.~(\ref{eq:diff1}) for one-nucleon losses in the limit $\Gamma^{\rm CEL}_{A,i}=0$ and $Q_{A,i}=0$.
In fact, the solution of a more general interaction network with generalized interaction rates $\Gamma_{(A,i)\to(B,j)}$ of the form~(\ref{eq:rates}) can be written
\begin{equation}\label{eq:fullsol}
N_{A,i}(t) = \sum_{j\geq i, B\geq A}\sum_{\mathbf{c}}\left(\prod_{l=1}^{n_c-1}\Gamma_{c_l\to c_{l+1}}\right)\sum_{k=1}^{n_c}\left[N_{B,j}(0)e^{-t\Gamma^{\rm tot}_{c_k}}+\int_0^t{\rm d}t'Q_{B,j}(t')e^{-(t-t')\Gamma^{\rm tot}_{c_k}}\right]\prod_{p=1(\neq k)}^{n_c}\frac{1}{\Gamma^{\rm tot}_{c_p}-\Gamma^{\rm tot}_{c_k}}\,,
\end{equation}
where we sum over all possible production chains $\mathbf{c}=\langle c_1,\ldots,c_{n_c}\rangle$ with intermediate nuclei of mass number $C$ in the energy bin $k$ --  denoted by the doublet $c_i = (C,k)$ -- and fixed endpoints $c_1 = (B,j)$ and $c_{n_c} = (A,i)$. The partial width $\Gamma_{c_l\to c_{l+1}}$ includes nucleon-disintegration ($\Gamma_{(A,i)\to (B,i)}$) as well as CEL ($\Gamma_{(A,i)\to (A,i-1)}$). A proof of Eq.~(\ref{eq:fullsol}) is given in Appendix~\ref{sec:appII}.

\begin{figure}[t]
\centering
\includegraphics[height=2in]{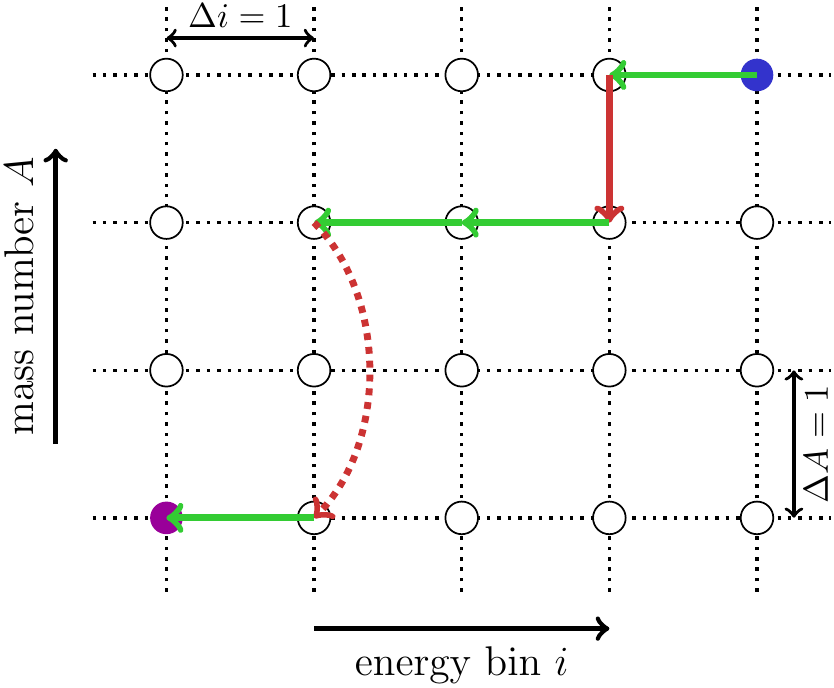}
\caption[]{A possible transition chain $\mathbf{c}$ between an initial configuration (blue dot) and a final configuration (magenta dot) including one-nucleon losses (red arrows), two-nucleon losses (red dotted arrows) and continuous energy loss (green arrows). For the exact analytic solution~(\ref{eq:fullsol}) all possible transition chains of this type are taken into account.}
\label{fig:pathexample}
\end{figure}

We can visualize the production chains $\mathbf{c}$ diagrammatically as {\it paths} along the configuration grid of nuclei, as shown in Fig.~\ref{fig:pathexample}. A horizontal link corresponds to a CEL transition whereas a vertical link denotes photo-disintegration. The color coding in Fig.~\ref{fig:pathexample} indicates the type of the transition $c_l\to c_{l+1}$ -- green for CEL, red for one-nucleon losses and red-dotted for two-nucleon losses\footnote{We will use later on ``generalized'' chains, where the transition $c_l\to c_{l+1}$ is not necessarily equal to $\Gamma_{c_l\to c_{l+1}}$.}. It is convenient to use this graphical representation as a short-hand notation for the terms of Eq.~(\ref{eq:fullsol}). To see this, we can write Eq.~(\ref{eq:fullsol}) in the form
\begin{equation}\label{eq:Green}
N_{A,i}(t) = \int_{0}^\infty{\rm d}t'\!\!\!\sum_{j\geq i, B\geq A}\!\!\!G(A,i,B,j;t-t')\left[Q_{B,j}(t')+\delta(t')N_{B,j}(0)\right]\,,
\end{equation}
where we define a Green's function $G(A,i,B,j;\Delta t)=\Theta(\Delta t)\sum_{\mathbf{c}}G(\mathbf{c};\Delta t)$ as a sum over the contribution {\it per path},
\begin{equation}
G(\mathbf{c};\Delta t)\equiv\left(\prod_{l=1}^{n_c-1}\Gamma_{c_l\to c_{l+1}}\right)\sum_{k=1}^{n_c}e^{-\Delta t\Gamma^{\rm tot}_{c_k}}\prod_{p=1(\neq k)}^{n_c}\frac{1}{\Gamma^{\rm tot}_{c_p}-\Gamma^{\rm tot}_{c_k}}\,.
\end{equation}
Each term $G(\mathbf{c};\Delta t)$ in the previous equation corresponds to a production chain on the configuration grid. We will use this graphical representation later for a perturbative expansions of Eq.~(\ref{eq:fullsol}).  

The interaction rates $\Gamma$ are not constant as the Universe expands. For example, the photo-disintegration rate with the CMB photons scales with red-shift as $\Gamma_A(z,E) = (1+z)^3\Gamma_A(0,(1+z)E)$, which follows from the adiabatic expansion of the CMB. Also, the nucleus emission rates $\mathcal{L}_A$ are not in general constant with time. A standard approach approximates the scaling with red-shift as a simple power-law over a finite red-shift distance, {\it e.g.}
\begin{equation}
\mathcal{L}_A(z,E) \equiv \Theta(z-z_{\rm min})\Theta(z_{\rm max}-z)(1+z)^n\mathcal{L}_A(0,E)\,.
\end{equation} 
We can account for the red-shift dependence of $\Gamma$ and $Q$ by summing Eqs.~(\ref{eq:fullsol}) over sufficiently small red-shift intervals, in which these quantities can be regarded as constant. Typically, intervals of $\Delta z\simeq0.01$ are sufficient for this approach.

Though the expression~(\ref{eq:fullsol}) is an {\it exact} analytical solution of the system of differential equations~(\ref{eq:diff1}), its calculation involves a large number of possible production chains and becomes numerically inefficient for large configuration grids.\footnote{In general, the numerical evaluation of expression (\ref{eq:fullsol}) requires a high computational precision. We use the publicly available multiple precision libraries {\tt GMP}~\cite{GMP} and {\tt MPFR}~\cite{MPFR} for this purpose.} For instance, for one-nucleon {\it and} two-nucleon losses the number of possible chains $F_{\Delta A}$ between nuclei with mass number $A$ and $B=A+\Delta A$ can be derived iteratively from the identity $F_{\Delta A+2}=F_{\Delta A}+F_{\Delta A+1}$ with $F_0=F_1=1$, which we recognize as the sequence of {\it Fibonacci} numbers. Hence, the total sum over different chains and $N$ primary nuclei in expression~(\ref{eq:fullsol}) involves $F_0+F_1+\ldots+F_{N-1} = F_{N+1}-1$ number of terms, which is a number that scales exponentially with $N$. Hence, considering all transitions via one-nucleon and two-nucleon losses between, say,  proton ($A=1$) and iron ($A=56$) becomes numerically very expensive even without considering transitions via CEL.

We show in the following that the exact expression~(\ref{eq:fullsol}) can be well approximated by the dominant production chain through one-nucleon losses. Corrections via two-nucleon losses and CEL can be treated perturbatively. As means of a comparative check, we obtain results using our analytic description, assuming a source  injection spectrum of  the form 
\begin{eqnarray}
J_{\rm inj} \propto E^{-\gamma}e^{-E/E_{\rm max}}.
\end{eqnarray}
These analytic results are compared against those obtained numerically through a Runge-Kutta method~\cite{GSL}.

\section{Perturbative Approach}\label{sec:perturbation}

The dominant contribution to the nucleon transitions in the CRB comes form one-nucleon losses with transition rate $\Gamma^{\rm 1N}_{A,i}$. In the following we focus on perturbative corrections to this dominant decay route from the contributions of two-nucleon losses and CEL with transition rates $\Gamma^{\rm 2N}_{A,i}$ and $\Gamma^{\rm CEL}_{A,i}$, respectively. 

\subsection{Two-Nucleon Losses}

We start with perturbative corrections from two-nucleon losses and assume, for the moment, that $\Gamma^{\rm tot}_{A,i} = \Gamma^{\rm 1N}_{A,i}+\Gamma^{\rm 2N}_{A,i}$ and $\Gamma^{\rm CEL}_{A,i}=0$. 
For a perturbative expansion it is convenient to rewrite Eq.~(\ref{eq:fullsol2}) as
\begin{equation}\label{eq:expansion}
N_{A,i}(t) = \sum_{B\geq A}\sum_{C=A}^B\mathcal{F}^{\,i}_{ABC}\left(\prod_{D=A+1}^B\!\!\!\!\Gamma^{\rm tot}_{D,i}\right)\left[N_{B,i}(0)e^{-t\Gamma^{\rm tot}_{C,i}}+\int_0^t{\rm d}t'Q_{B,i}(t')e^{-(t-t')\Gamma^{\rm tot}_{C,i}}\right]\!\!\prod_{D=A(\neq C)}^B\frac{1}{\Gamma^{\rm tot}_{D,i}-\Gamma^{\rm tot}_{C,i}}\,,
\end{equation}
with
\begin{equation}\label{eq:Fkij}
\mathcal{F}^{\,i}_{ABC} \equiv \sum_{\mathbf{c}}\left(\prod_{l=1}^{n_c-1}\frac{\Gamma_{c_{l}\to c_{l+1}}}{\Gamma^{\rm tot}_{c_{l}}}\right)\prod_{D=A+1(\notin\mathbf{c})}^{B}\left(1-\frac{\Gamma^{\rm tot}_{C,i}}{\Gamma^{\rm tot}_{D,i}}\right)\,.
\end{equation}
We can define a perturbative expansion of Eq.~(\ref{eq:Fkij}) in terms of sub-dominant branching ratios of two-nucleon production, $\Gamma^{\rm 2N}_{C,i}/\Gamma^{\rm tot}_{C,i}$. The leading order (LO) contribution, $\mathcal{F}^{\,i,{\rm LO}}_{ABC}=1$, reproduces the approximation of Ref.~\cite{Hooper:2008pm}. The next-to-leading order (NLO) contribution can be written as
\begin{equation}
\mathcal{F}^{\,i, {\rm NLO}}_{ABC}=  \sum_{D=A+2}^{B}\frac{\Gamma^{\rm 2N}_{D,i}}{\Gamma^{\rm tot}_{D,i}}\left(1-\frac{\Gamma^{\rm tot}_{C,i}}{\Gamma^{\rm tot}_{D-1,i}}\right)- \sum_{D=A+1}^{B}\frac{\Gamma^{\rm 2N}_{D,i}}{\Gamma^{\rm tot}_{D,i}} \,.
\end{equation}

\begin{figure}[t]
\centering
\includegraphics[height=1.65in]{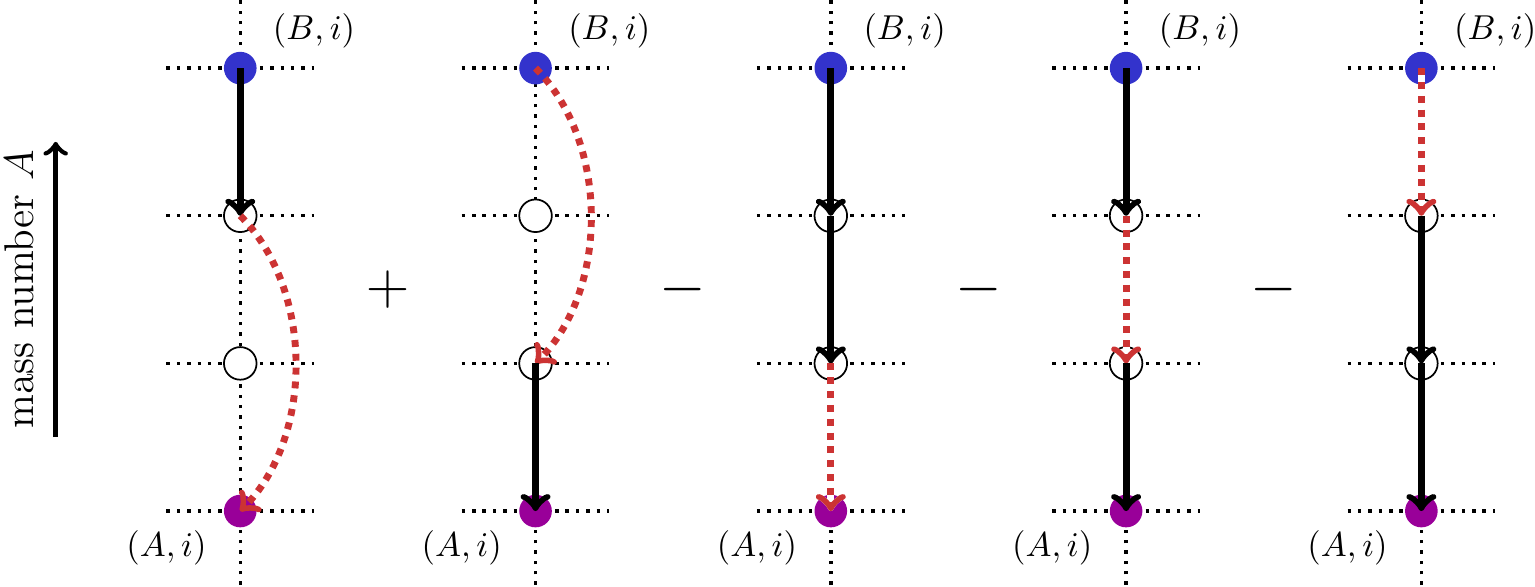}
\caption[]{A graphic representation of the NLO paths contributing in the first correction $N^{(1)}_{A,i}$ for $\Delta A = 3$ (see Eq.~(\ref{eq:NLO2N})). The black arrows indicate transitions between configurations with total transition rate $\Gamma^{\rm tot}_{A,i}=\Gamma^{\rm 1N}_{A,i}+\Gamma^{\rm 2N}_{A,i}$ and the dotted red arrows two-nucleon transition rates $\Gamma^{\rm 2N}_{A,i}$, respectively. Note that these types of graphs contribute with {\it opposite} sign in Eq.~(\ref{eq:NLO2N}).}
\label{fig:NLO2N}
\end{figure}

We can most easily visualize these terms by a perturbative expansion of the nucleon densities, 
\begin{equation}\label{eq:expansion2}
N_{A,i} = \sum_{n\geq0} N^{(n)}_{A,i}\,,
\end{equation} 
where the terms $N^{(n)}_{A,i}$ are solutions to the set of differential equations,
\begin{align}\label{eq:diff2}
\dot N^{(0)}_{A,i} &= - \Gamma^{\rm tot}_{A,i} N^{(0)}_{A,i}+ \Gamma^{\rm tot}_{A+1,i} N^{(0)}_{A+1,i} + Q_{A,i}\,,\\
\dot N^{(n)}_{A,i} &= - \Gamma^{\rm tot}_{A,i} N^{(n)}_{A,i}+ \Gamma^{\rm tot}_{A+1,i} N^{(n)}_{A+1,i}+\Gamma^{\rm 2N}_{A+2,i}N^{(n-1)}_{A+2,i}-\Gamma^{\rm 2N}_{A+1,i}N^{(n-1)}_{A+1,i} \qquad(n>0)\,.\nonumber
\end{align}
For the moment, we assume that the total width $\Gamma^{\rm tot}_{A,i}$ is the sum of one-nucleon and two-nucleon losses. This, however, can be generalized to the total photo-disintegration rate for general nucleon losses (see section~\ref{sec:generalloss}). As an initial condition we define $N^{(n)}_{A,i}(0)=0$ for $n>0$ and $N^{(0)}_{A,i}(0)=N_{A,i}(0)$. Note, that with this initial condition the expansion (\ref{eq:expansion2}) becomes finite and hence converges trivially. Each term $N^{(n)}$ corresponds, by construction, to the $n$-th order correction of $\mathcal{F}$. We can write the NLO correction explicitly as
\begin{equation}\label{eq:NLO2N}
N^{(1)}_{A,i}(t) = \sum_{B\geq A}\sum_{C=A}^B\left(\prod_{D=A+1}^B\!\!\!\!\Gamma^{\rm tot}_{D,i}\right)\left[\int_0^t{\rm d}t'\left(\Gamma^{\rm 2N}_{B+2,i}N^{(0)}_{B+2,i}(t')-\Gamma^{\rm 2N}_{B+1,i}N^{(0)}_{B+1,i}(t')\right)e^{-(t-t')\Gamma^{\rm tot}_{C,i}}\right]\!\!\prod_{D=A(\neq C)}^B\frac{1}{\Gamma^{\rm tot}_{D,i}-\Gamma^{\rm tot}_{C,i}}\,.
\end{equation}
Inserting the LO solution in Eq.~(\ref{eq:NLO2N}) and following similar algebraic steps as in Appendix~\ref{sec:appII} one can identify $N^{(1)}$ as the {\it difference} of contributions form paths $\langle(A,i),\ldots,(B,i)\rangle$ with length $B-A+1$ and $B-A$, respectively, with the {\it single} insertion of a two-nucleon loss step into the decay chain. This is displayed diagrammatically in Fig.~\ref{fig:NLO2N} for the case $\Delta A=3$. 

\begin{figure}[t]
\centering
\includegraphics[width=0.7\linewidth]{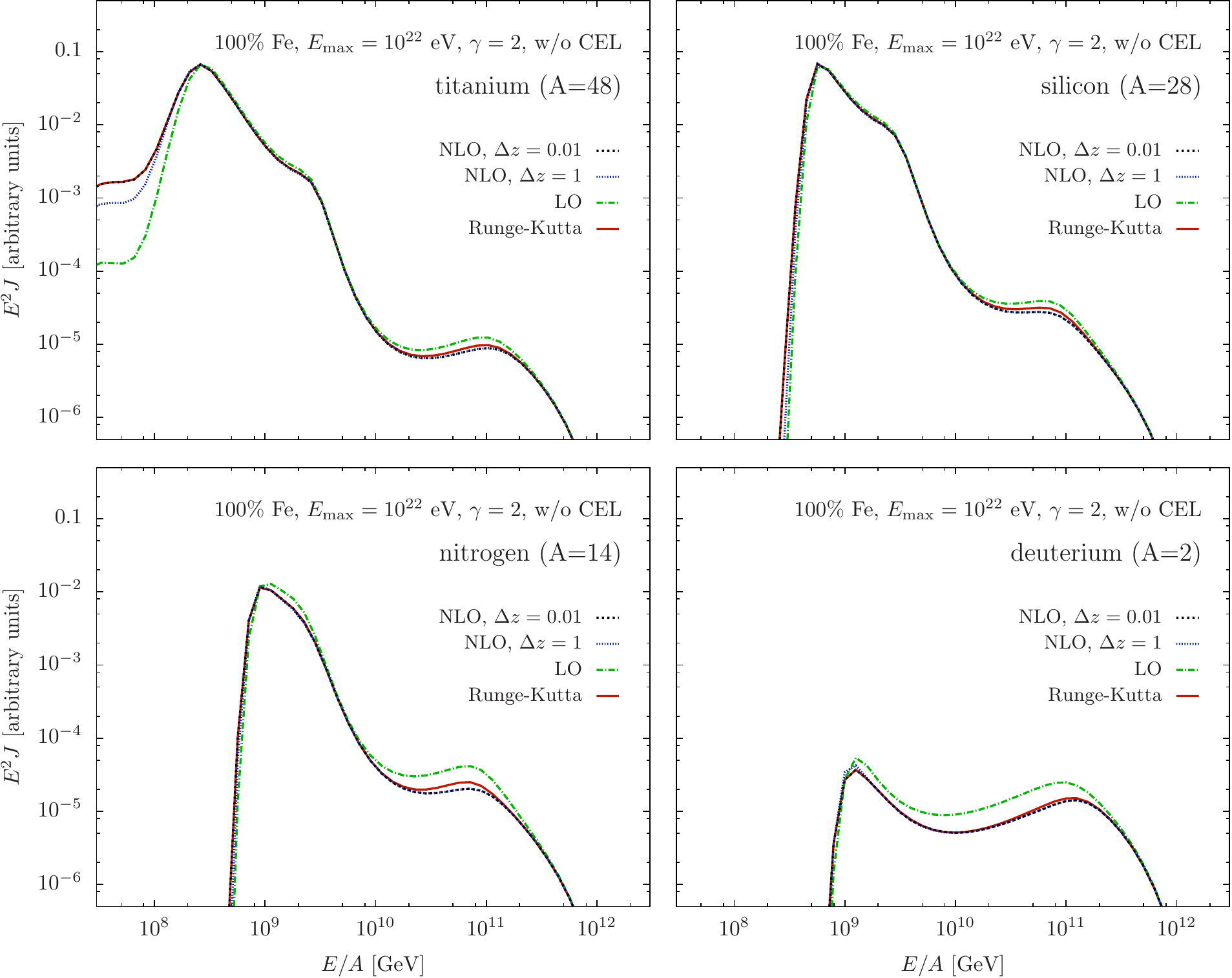}
\caption[]{The solution (\ref{eq:expansion}) at leading-order (LO) and up to next-to-leading order (NLO) (Eq.~(\ref{eq:NLO2N})) for one-nucleon (1N) and two-nucleon (2N) losses. To aid the comparison between the results, we ignore the evolution of the nucleon emission rates and interaction rates with red-shift and sum over red-shift steps $\Delta z =0.01$. We compare the LO and NLO analytic results to a numerical solution via a Runge-Kutta method~\cite{GSL}.}
\label{fig:samples2N}
\end{figure}

Note, that we can also express Eqs.~(\ref{eq:NLO2N}) as a matrix equation of the form,
\begin{equation}\label{eq:NLO2N2}
N^{(1)}_{A,i}(\Delta t)\simeq \sum_{B\geq A}\left(\mathcal{X}_{AB,i}(\Delta t)N_{B,i}(0)+\mathcal{Y}_{AB,i}(\Delta t)Q_{B,i}\right)\,.
\end{equation}
The matrices $\mathcal{X}(\Delta t)$ and $\mathcal{Y}(\Delta t)$ are in general only slowly changing with the red-shift scaling of the background radiation. It is hence possible to improve the NLO results by introducing sufficiently small time intervals $\Delta z$ and apply Eq.~(\ref{eq:NLO2N2}) repeatedly. 

We show the LO and NLO results of our approach in comparison to a numerical solution via a Runge-Kutta method in Fig.~\ref{fig:samples2N}. For simplicity, we assume that CEL is absent and that source terms and interaction rates are constant throughout the integration domain $0<z<1$. The NLO contributions are shown for two cases. In the case ``$\Delta z=1$'' we calculate the NLO contribution directly by Eq.~(\ref{eq:NLO2N}). The case ``$\Delta z = 0.01$'' shows the improvement of the NLO contribution by a repeated application of Eq.~(\ref{eq:NLO2N2}) for the corresponding time interval - 100 times in this case. In most cases, the LO approximation is already satisfactory~\cite{Hooper:2008pm}.

\subsection{General Photo-Disintegration Losses}\label{sec:generalloss}

For high mass nuclei ($A\gtrsim40$) of the PSB-chain one-nucleon and two-nucleon losses constitute more than 90\% of the total photo-disintegration rate as can be seen in the Table~\ref{tab:PSB}. However, for low mass nuclei the emission of $\alpha$ particles (as well as deuterons (D) and tritons (T)) can become important. As in the previous section, we can organize these sub-leading contributions via the expansion~(\ref{eq:expansion2}). For instance, the additional contribution from $\alpha$ particle loss can be introduced at NLO ($n>0$) as
\begin{align}\label{eq:NLOalpha}
\dot N^{(n)}_{A,i} = - \Gamma^{\rm tot}_{A,i} N^{(n)}_{A,i}+ \Gamma^{\rm tot}_{A+1,i} N^{(n)}_{A+1,i}+\Gamma^{\rm 2N}_{A+2,i}N^{(n-1)}_{A+2,i}-\Gamma^{\rm 2N}_{A+1,i}N^{(n-1)}_{A+1,i}+\Gamma^{\alpha}_{A+4,i}N^{(n-2)}_{A+4,i}-\Gamma^{\alpha}_{A+1,i}N^{(n-2)}_{A+1,i}\,,
\end{align}
where we now have to include $\alpha$ emission in the definition of the total rate, $\Gamma^{\rm tot}_{A,i} = \Gamma^{\rm 1N}_{A,i}+\Gamma^{\rm 2N}_{A,i}+\Gamma^{\alpha}_{A,i}$. The treatment of these additional photo-disintegration channels is completely analogous to the case of two-nucleon losses.

\subsection{Continuous Energy Losses}

We next consider the contribution of CEL to the solution~(\ref{eq:fullsol}). In this case we have to include all possible paths in Eq.~(\ref{eq:fullsol}) that allow for both, variation of energy {\it and} mass number as the one shown in the left panel of Fig.~\ref{fig:pathexample}. Similar to the discussion of two-nucleon losses, the number of possible paths becomes very large. For the remainder of this section we consider only one-nucleon photo-disintegration losses together with CEL and, hence, $\Gamma^{\rm tot}_{A,i} = \Gamma^{\rm 1N}_{A,i}+\Gamma^{\rm CEL}_{A,i}$. Despite this simplification there are still $(\Delta A+\Delta i)!/(\Delta A)!/(\Delta i)!$ different paths in total between the two configurations $(A,i)$ and $(A+\Delta A,i+\Delta i)$. This becomes computationally very expensive for long production chains, as we already observed for the introduction of two-nucleon losses. 

We can account for CEL transitions as effective source terms in the differential equations (\ref{eq:diff1}). This turns out to be an efficient way for determining the resulting spectra. As before, we can use the perturbative expansion (\ref{eq:expansion2}) of the nucleon densities, where the terms $N^{(n)}_{A,i}$ are now solutions to the set of differential equations,
\begin{align}\label{eq:diff3}
\dot N^{(0)}_{A,i} &= - \Gamma^{\rm tot}_{A,i} N^{(0)}_{A,i}+ \Gamma^{\rm tot}_{A+1,i} N^{(0)}_{A+1,i} + Q_{A,i}\,,\\
\dot N^{(n)}_{A,i} &= - \Gamma^{\rm tot}_{A,i} N^{(n)}_{A,i}+ \Gamma^{\rm tot}_{A+1,i} N^{(n)}_{A+1,i}+\Gamma^{\rm CEL}_{A,i+1}N^{(n-1)}_{A,i+1}-\Gamma^{\rm CEL}_{A+1,i}N^{(n-1)}_{A+1,i} \qquad(n>0)\,.\nonumber
\end{align}
Here, the total width $\Gamma^{\rm tot}_{A,i}=\Gamma^{\rm 1N}_{A,i}+\Gamma^{\rm CEL}_{A,i}$ is now for the sum of one-nucleon and CEL, though in general it would receive contributions from all exclusive channels. Note that with the initial condition $N^{(n)}_{A,i}(0)=0$ for $n>0$ the expansion (\ref{eq:expansion2}) of $N_{A,i}$ is finite if a finite set of energy bins and nuclei is considered, $A\leq A_{\rm max}$ and $i\leq i_{\rm max}$. More specifically, the expansion of $N_{A,i}$ only includes non-zero terms $N^{(n)}_{A,i}$ for $n\leq(A_{\rm max}+i_{\rm max})-(A+i)$.

\begin{figure}[t]
\centering
\includegraphics[height=1.65in]{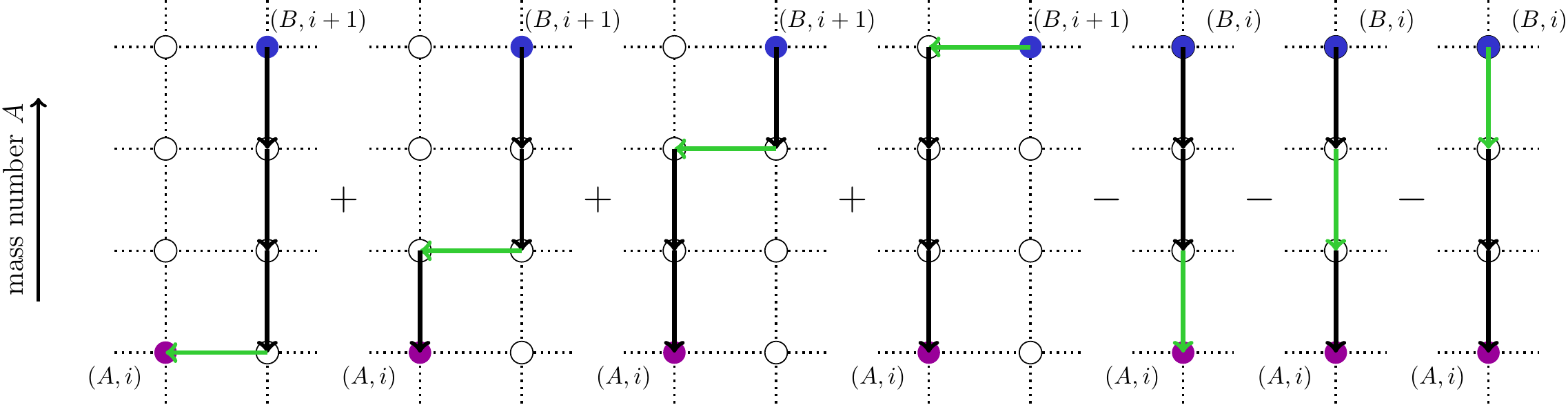}
\caption[]{A graphic representation of the NLO paths contributing in the first correction $N^{(1)}_{A,i}$ for $\Delta A = 3$ (see Eq.~(\ref{eq:NLOCEL})). The black arrows indicate transitions between configurations with total transition rate $\Gamma^{\rm tot}_{A,i}=\Gamma^{\rm 1N}_{A,i}+\Gamma^{\rm CEL}_{A,i}$ and the green arrows transitions with CEL rate $\Gamma^{\rm CEL}_{A,i}$, respectively. Note that these types of graphs contribute with {\it opposite} sign in Eq.~(\ref{eq:NLOCEL}).}
\label{fig:NLOCEL}
\end{figure}

The first term $N^{(0)}_{A,i}$ in the expansion of $N_{A,i}$ is our familiar solution for the one-nucleon loss case (\ref{eq:expansion}) where the partial width is replaced by the total width. The second term $N^{(1)}$ can be evaluated explicitly by an insertion of $N^{(0)}$,
\begin{equation}\label{eq:NLOCEL}
N^{(1)}_{A,i}(t) = \sum_{B\geq A}\sum_{C=A}^B\left(\prod_{D=A+1}^B\!\!\!\!\Gamma^{\rm tot}_{D,i}\right)\left[\int_0^t{\rm d}t'\left(\Gamma^{\rm CEL}_{B,i+1}N^{(0)}_{B,i+1}(t')-\Gamma^{\rm CEL}_{B+1,i}N^{(0)}_{B+1,i}(t')\right)e^{-(t-t')\Gamma^{\rm tot}_{C,i}}\right]\!\!\prod_{D=A(\neq C)}^B\frac{1}{\Gamma^{\rm tot}_{D,i}-\Gamma^{\rm tot}_{C,i}}\,.
\end{equation}
After some algebraic manipulations one can identify $N^{(1)}$ as the {\it difference} of contributions form paths $\langle(A,i),\ldots,(B,i+1)\rangle$ and $\langle(A,i),\ldots,(B,i)\rangle$ with the {\it single} insertion of CEL step into the decay chain. This is indicated diagrammatically in Fig.~\ref{fig:NLOCEL} for the case $\Delta A=3$. 

Note that the NLO correction for CEL only introduces transitions between the energy bins $i$ and $i+1$. Hence, the NLO  solution~(\ref{eq:NLOCEL}) can not be considered as a small correction to the full solution if the contribution from CEL becomes large, $\Delta t\Gamma^{\rm CEL}_{A,i}\gtrsim 1$. However, in analogy to the case of two-nucleon losses we can write the NLO contribution as a matrix equation
\begin{equation}\label{eq:NLOCEL2}
N^{(1)}_{A,i}(\Delta t)\simeq \sum_{B\geq A}\left(\mathcal{X}_{AB,i}(\Delta t)N_{B,i}(0)+\mathcal{Y}_{AB,i}(\Delta t)Q_{B,i}+\mathcal{V}_{AB,i}(\Delta t)N_{B,i+1}(0)+\mathcal{W}_{AB,i}(\Delta t)Q_{B,i+1}\right)\,.
\end{equation}
If we consider {\it sufficiently} small time intervals $\Delta t$ such that $\Delta t\Gamma^{\rm CEL}_{A,i}\ll1$ we can approximate the exact solution by a repeated application of Eq.~(\ref{eq:NLOCEL2}). The transition matrices $\mathcal{X}(\Delta t)$, $\mathcal{Y}(\Delta t)$, $\mathcal{V}(\Delta t)$ and $\mathcal{W}(\Delta t)$ are only slowly changing with the scaling of the background radiation. It is hence only necessary to re-evaluate these matrices on large time-scales; typically $\Delta z\simeq 0.01$ is sufficient for the propagation of heavy nuclei. Thus, results obtained by the application of this procedure should be considered {\it semi-analytic}.

\begin{figure}[ph!]
\centering
\includegraphics[width=0.7\linewidth]{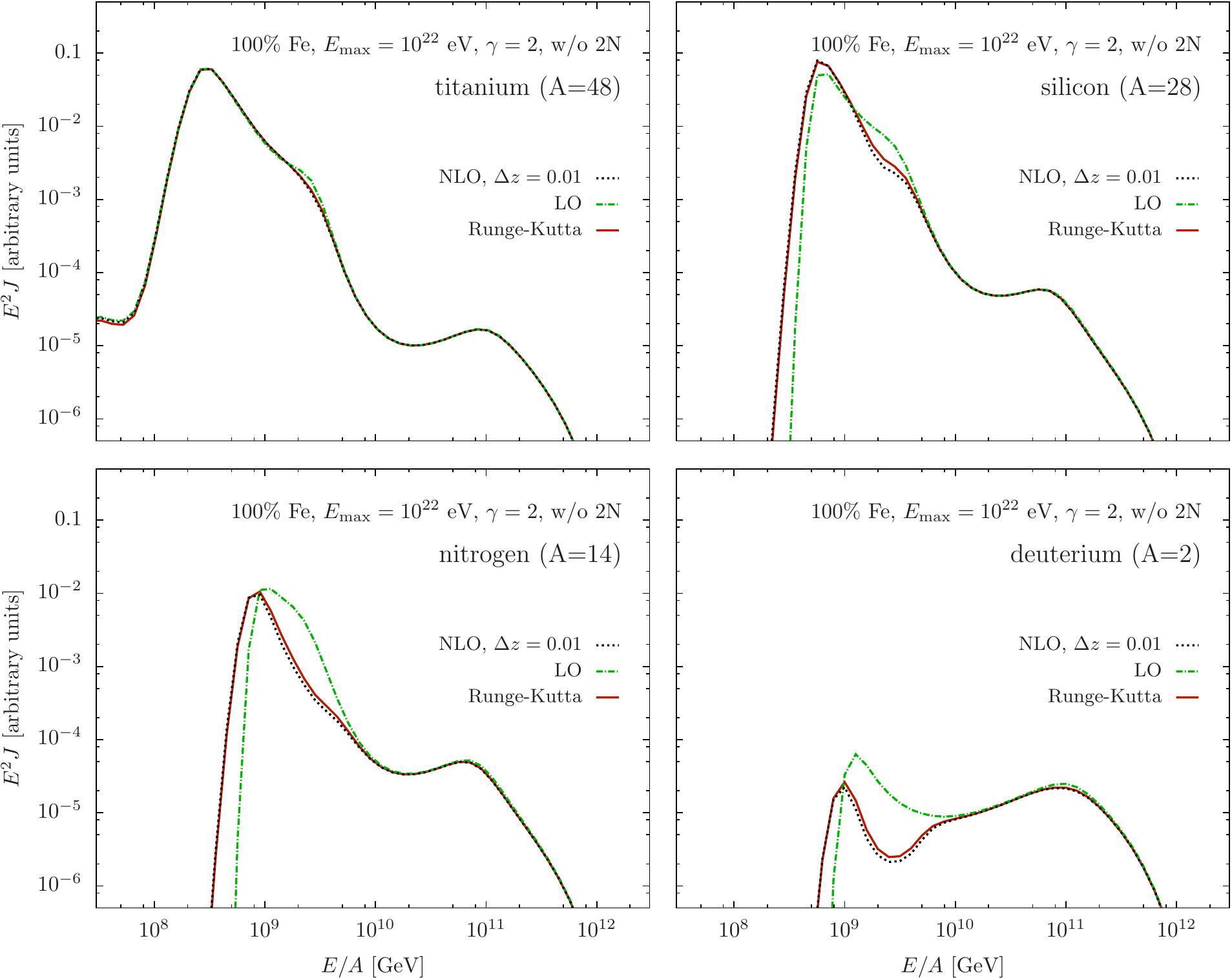}
\caption[]{Comparison of the terms in expression (\ref{eq:expansion}) up to next-to-leading order (NLO) with the numerical solution via a Runge-Kutta method including one-nucleon (1N) and continuous energy losses (CEL). To aid the comparison between the results, we ignore the evolution of the nucleon emission rates and interaction rates with red-shift and sum over red-shift steps $\Delta z =0.01$.}\label{fig:samplesCEL}
\vspace{0.2in}
\includegraphics[width=0.7\linewidth]{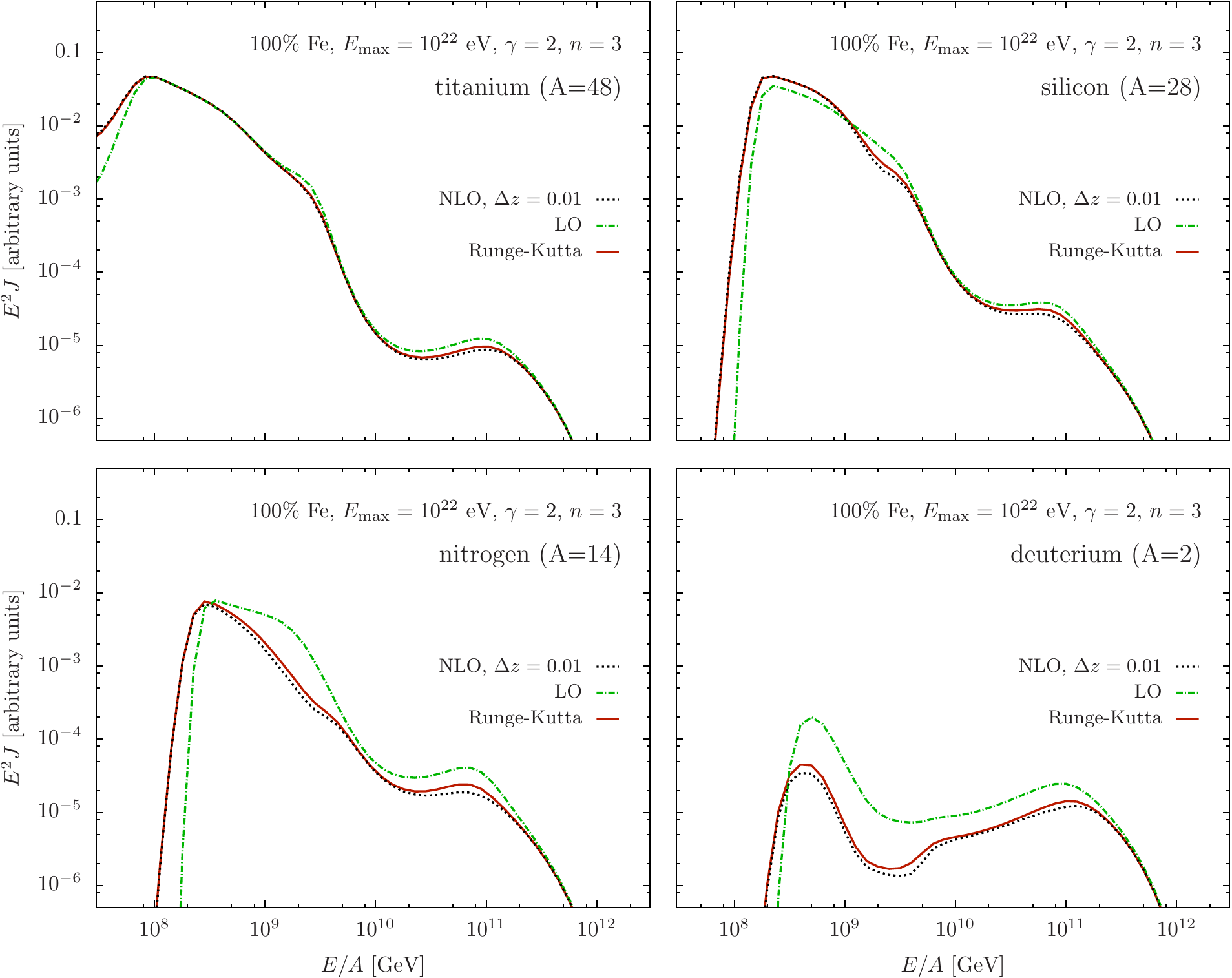}
\caption[]{
The full NLO correction for two-nucleon and continuous energy losses in comparison with the numerical solution. For this result we took into account source evolution and red-shift effects, assuming that the nucleon emission rates scale as $(1+z)^3$ and sum over red-shift steps $\Delta z =0.01$.}\label{fig:samplesALL}
\end{figure}

Figure~\ref{fig:samplesCEL} shows the results of the LO and NLO energy flux spectra compared with results obtained using a Runge-Kutta method. For simplicity, we again consider constant source terms and interaction rates and assume that two-nucleon losses are absent. The repeated application of Eq.~(\ref{eq:NLOCEL2}) reproduces the numerical solution well.  For heavy nuclei (and hence ``short'' transitions from primary iron) or large energies $E/A>10^{10}$~GeV the LO contribution is already an excellent approximation.

\subsection{Secondary Proton and Helium Spectra}

Finally, we discuss an expansion of the spectrum of primary and secondary protons\footnote{We do not distinguish between protons and neutrons in the following, assuming a prompt decay of secondary neutrons.} and helium. This case is slightly different from the propagation of heavy nuclei, since there are additional contributions from the channels ($\gamma$,N), ($\gamma$,2N), ($\gamma$,$\alpha$), ($\gamma$,N$\alpha$) and ($\gamma$,2$\alpha$).
Secondary nucleon production follows the differential equation 
\begin{equation}\label{eq:diff_proton}
\dot N_{1,i} \simeq \Gamma^{\rm CEL}_{1,i+1}N_{1,i+1}-\Gamma^{\rm CEL}_{1,i}N_{1,i} + \sum_{A\geq2}\Gamma^{\rm eff, N}_{A,i}N_{A,i} + Q_{1,i}\,, 
\end{equation}
where the effective nucleon production rate $\Gamma^{\rm eff, N}_{A,i}$ from transitions $(A,i)\to(1,i)$ is defined as 
\begin{equation}\label{eq:Geffp}
\Gamma^{\rm eff, N}_{A,i} \equiv \Gamma^{\rm 1N}_{A,i}  + 2\Gamma^{\rm 2N}_{A,i} + \Gamma^{\rm N\alpha}_{A,i}+\delta_{A2}\Gamma^{\rm 1N}_{2,i}+\delta_{A3}\Gamma^{\rm 2N}_{3,i}\,(+\delta_{A6}\Gamma^{\rm N\alpha}_{6,i})\,,
\end{equation}
with $\delta_{AB} = 1$ if $A=B$ and zero otherwise\footnote{For $N_{A,i}' = N_{A,i}/2$ for $A=2,3,4$ we re-define ${\Gamma^{\rm eff}_{A,i}}'=2\Gamma^{\rm eff}_{A,i}$.}. Note, that the last term in (\ref{eq:Geffp}) is assumed absent in the PSB-chain. Photo-hadronic interactions of the protons can be determined using the Monte Carlo Package {\tt SOPHIA}~\cite{Mucke:1999yb}. Here, we approximate photo-pion interactions of the protons as a continuous energy loss process in addition to Bethe-Heitler pair production. The differential equation~(\ref{eq:diff_proton}) is of the same form as Eq.~(\ref{eq:diff1}) and we can hence write its exact solution in the form (\ref{eq:fullsol}).

\begin{figure}[t]
\centering
\includegraphics[height=1.7in]{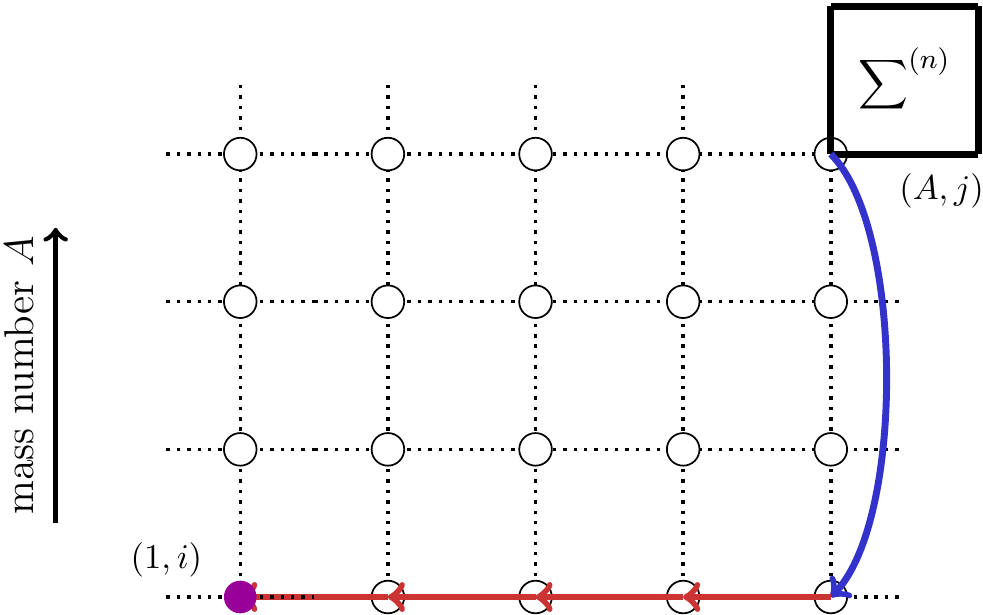}
\caption[]{An example of an n-th order production path contributing to $N^{(n)}_{1,i}$ including the effective nucleon production rate $\Gamma^{\rm eff}_{B,j}$ (blue arrow). The box in the top right corners indicate the {\it complete} sum over all possible $n$-th order contributions ``${\sum}^{(n)}$'' to the production chain of the nucleon $(B,j)$ with $B\geq 2$ and $j\geq i$. The red arrows indicate the CEL contribution for protons, including energy loss by Bethe-Heitler pair production and photo-pion production.}
\label{fig:protonNLO}
\end{figure}

With the expansion (\ref{eq:expansion}) we can write the set of evolution equations as
\begin{align}\label{eq:diff_proton2}
\dot N^{(0)}_{1,i} &\simeq -\Gamma^{\rm CEL}_{1,i}N^{(0)}_{1,i}+\Gamma^{\rm CEL}_{1,i+1}N^{(0)}_{1,i+1}+\sum_{A\geq2}\Gamma^{\rm eff}_{A,i}N^{(0)}_{A,i} + Q_{1,i}\,,\\
\dot N^{(n)}_{1,i} &\simeq -\Gamma^{\rm CEL}_{1,i}N^{(n)}_{1,i}+\Gamma^{\rm CEL}_{1,i+1}N^{(n)}_{1,i+1}+\sum_{A\geq2}\Gamma^{\rm eff}_{A,i}N^{(n)}_{A,i}
\qquad(n>0)\,.\nonumber
\end{align}
In contrast to the case of nuclei, we cannot treat CEL of the protons as a second order effect. Nevertheless, with the set of differential equations~(\ref{eq:diff_proton2}) and the boundary condition $N^{(0)}_{1,i}(0)=N_{1,i}(0)$ and $N^{(n)}_{1,i}(0) =0$ for $n>0$ the expansion (\ref{eq:expansion}) is finite since the expansion of $N_{A,i}$ is finite. Explicitly, we can write the n-th order contribution as
\begin{equation}\label{eq:NTHORDER}
N^{(n)}_{1,i}(t) = \sum_{B\geq 2}\sum_{j\geq i}\left(\prod_{k=i+1}^j\!\!\!\!\Gamma^{\rm CEL}_{1,k}\right)\sum_{k=i}^j\left[\int_0^t{\rm d}t'\left(\Gamma^{\rm eff}_{B,j}N^{(n)}_{B,j}(t')\right)e^{-(t-t')\Gamma^{\rm CEL}_{1,k}}\right]\!\!\prod_{l=i(\neq k)}^j\frac{1}{\Gamma^{\rm CEL}_{1,l}-\Gamma^{\rm CEL}_{1,k}}\,.
\end{equation}
Again, these contributions to the proton spectra can be expressed via diagrams indicated in Fig.~\ref{fig:protonNLO}. By definition, the term $N_{1,i}^{(n)}$ depend on all possible n-th order production chains of intermediate nuclei $(B,j)$, that are indicated as the boxes in the top right corner of the diagrams. 

Similarly, the emission of $\alpha$ particles in the channels ($\gamma$,$\alpha$), ($\gamma$,2$\alpha$) and ($\gamma$,N$\alpha$) rate can be described by the differential equation
\begin{equation}\label{eq:diff_helium}
\dot N_{4,i} \simeq  \Gamma^{\rm CEL}_{4,i+1}N_{4,i+1}-\Gamma^{\rm CEL}_{4,i}N_{4,i} -(\Gamma^{\rm 1N}_{4,i}+\Gamma^{\rm 2N}_{4,i})N_{4,i} + \sum_{A\geq2}\Gamma^{\rm eff,\alpha}_{A,i}N_{A,i} + Q_{4,i}\,, 
\end{equation}
with an effective production rate 
\begin{equation}\label{eq:Geffalpha}
\Gamma^{\rm eff, \alpha}_{A,i} \equiv \Gamma^{\rm \alpha}_{A,i} + \Gamma^{\rm N\alpha}_{A,i} + 2\Gamma^{2\alpha}_{A,i} +\delta_{A12}\Gamma^{\rm 2\alpha}_{12,i}+\delta_{A9}\Gamma^{\rm N\alpha}_{9,i}\,(+\delta_{A8}\Gamma^{\rm \alpha}_{8,i})\,.
\end{equation}
Again, the last term in~(\ref{eq:Geffalpha}) is absent in the PSB-chain considered in our calculation. In principle, we can treat these contributions analogously to the case of the protons. However, the relative contribution from $\alpha$ particle emission is only small if we consider heavy primary nuclei like ${}^{56}$Fe and can be neglected in this case.

\begin{figure}[t]
\centering
\includegraphics[height=0.33\linewidth]{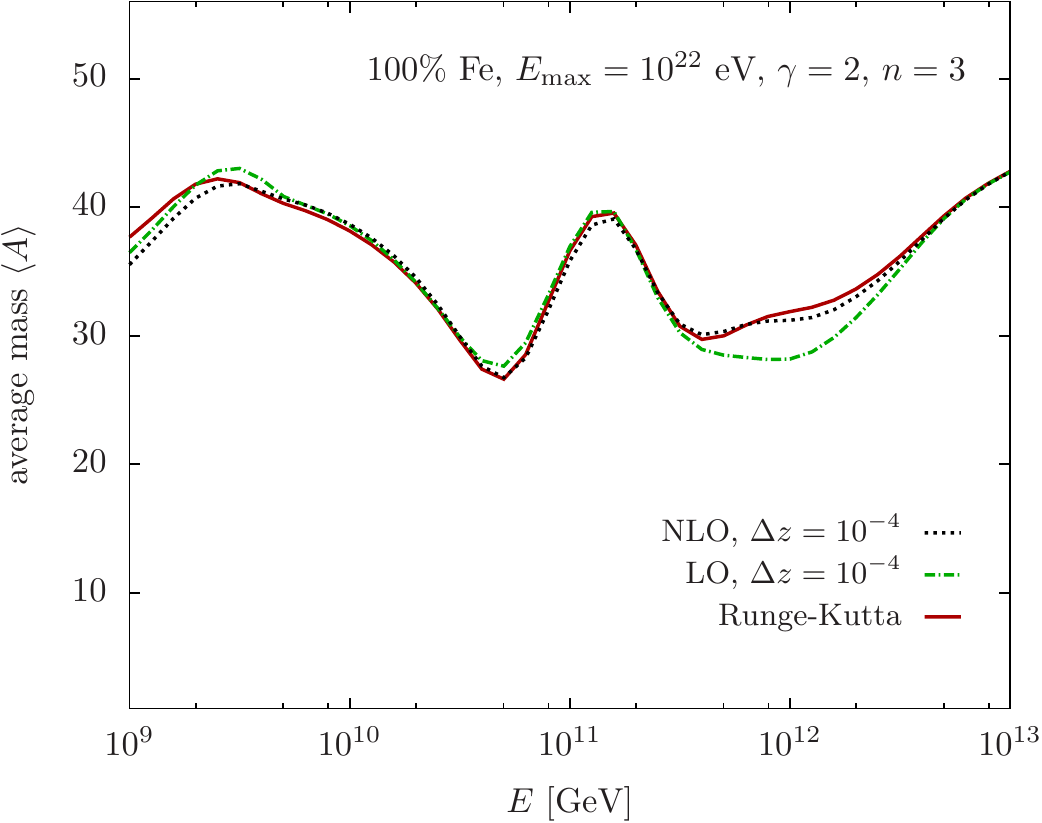}\hspace{0.3in}\includegraphics[height=0.33\linewidth]{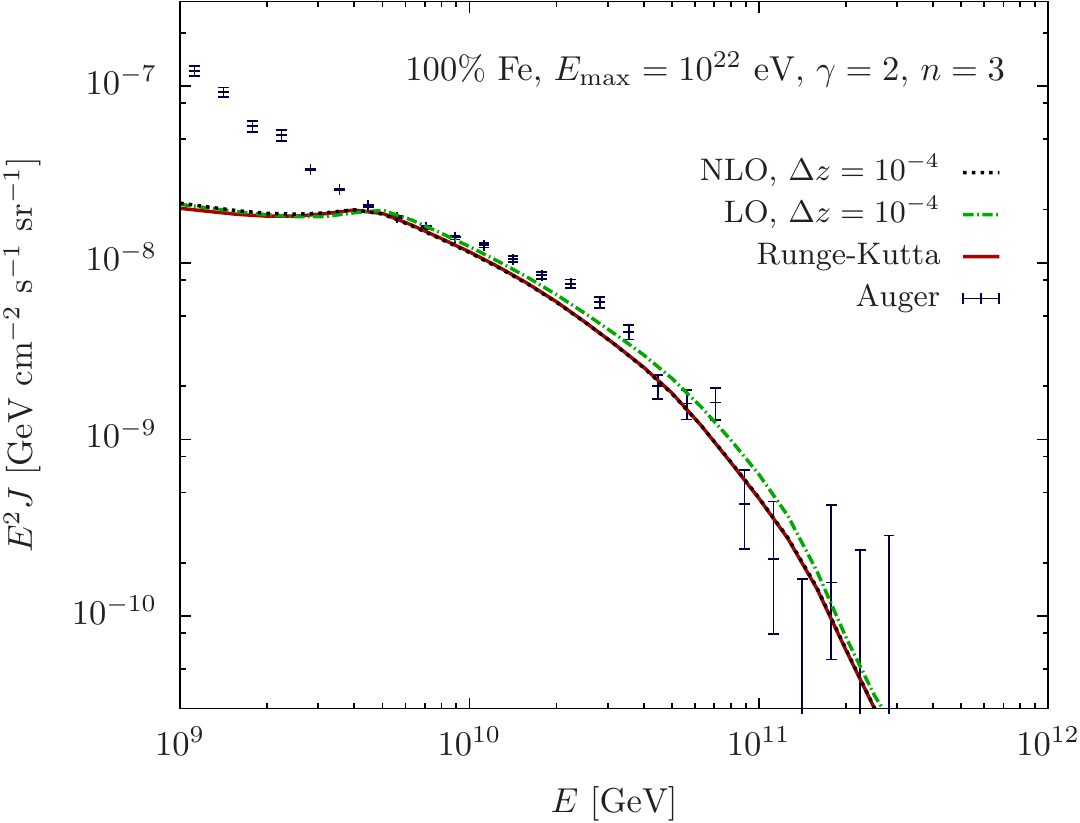}
\caption[]{{\bf Left:}  The average mass number from a pure-iron $E^{-2}$-flux with $E_{\rm max}=10^{22}$~eV and source evolution parameters $n=3$ and $z_{\rm max}=1$. We show the full numerical solution in comparison with the LO and NLO analytic equation including the exclusive channels $(\gamma,N)$, $(\gamma,2N)$, $(\gamma,\alpha)$, $(\gamma,N\alpha)$ and $(\gamma,2\alpha)$ as well as CEL. {\bf Right:} The total energy flux for the same parameters plotted against recent Auger measurements \cite{Abraham:2010mj}. The LO results are in good agreement with those shown in Fig.~4 of \cite{Hooper:2008pm}.}
\label{fig:averageA}
\end{figure}

The sum over diagrams of the type shown in Fig.~\ref{fig:protonNLO} involve a large number of intermediate configurations $(B,j)$ and the calculation can become time-consuming. For a more efficient calculation of the proton spectra we can utilize the total conservation of nucleons per energy bin within sufficiently small time-steps with $\Delta t\Gamma^{\rm CEL}_{1,i} \ll 1$. In this case the flux can be well approximated as
\begin{equation}\label{eq:NLPapprox}
N_{1,i}(\Delta t) \simeq N_{1,i}(0) +\Delta t Q_{1,i} +\Delta t \left[\Gamma^{\rm CEL}_{1,i+1}N_{1,i+1}(0)-\Gamma^{\rm CEL}_{1,i}N_{1,i}(0)\right]+\sum_{A\geq2}A\left[N_{A,i}(0)+\Delta t Q_{A,i}-N_{A,i}(\Delta t)\right]\,.
\end{equation}
With this approximation, and using the NLO contribution of the exclusive channels $(\gamma,N)$, $(\gamma,2N)$, $(\gamma,\alpha)$, $(\gamma,N\alpha)$ and $(\gamma,2\alpha)$ as well as CEL for the spectra of nuclei, we show in the left panel of Fig.~\ref{fig:averageA} the average mass number $\langle A\rangle$ in comparison with the analytic result. The right panel of Fig.~\ref{fig:averageA} shows the total energy flux of nuclei for the NLO analytic solution compared to the numerical result. For these results, time steps of $\Delta z=10^{-4}$ have been used in order for the proton contribution to the total flux to be calculated with the necessary accuracy. The LO approximation is already in excellent approximation to the data of CR observatories considering the large systematic and statistical uncertainties of the CR spectra and the average mass composition. All spectral features of the quantities and their overall scale are well reproduced by the LO contribution. Improvements to the LO result, however, are made by the NLO contributions, whose results leave only a very mild discrepancy with the Runge-Kutta results at energies below $10^{9.5}$~GeV.

\section{Conclusions}\label{sec:conclusions}

In this work we have developed further an analytic solution for the fluxes of UHE CR nuclei from extragalactic sources. We have shown that in most cases the spectra are well approximated by the analytic solution already given in Ref.~\cite{Hooper:2008pm}, which dealt with the dominant energy loss channel of single nucleon transitions between nuclei. We have here expanded on this approach through the introduction of NLO corrections from two-nucleon and CEL. The introduction of these terms was shown to further improve the accuracy of the analytic description. In order for these results to take into account the slow variation of interaction and emission rates with red-shift as well as CEL we incorporated our result into a semi-analytic framework. The semi-analytic results obtained were found to be in excellent agreement with results obtained through a purely numerical Runge-Kutta approach. 

The prospects of determining the nature of extragalactic UHE CRs and their sources in the near future are promising. Ongoing {\it direct} hybrid measurements of UHE CRs by the Auger collaboration continue, with the opportunity now existing for an independent verification of these results by other hybrid experiments such as the Telescope Array \cite{Tokuno:2010zz}. These measurements allow the possibility for a coherent picture of the UHE CR flux, composition, and arrival direction anisotropy to emerge. Present and ongoing {\it indirect} measurements of the secondary particles produced by UHE CRs during their acceleration and propagation are also capable of constraining the UHE CR composition and their sources. For instance, the simultaneous emission of neutrinos arising from proton-proton and/or proton-photon interactions in extra-galactic protons sources can serve as a test of low energy crossover scenarios~\cite{Berezinsky:2002nc} of extra-galactic protons~\cite{Ahlers:2005sn,Ahlers:2009rf}. Photo-pion interactions by extra-galactic protons in the CMB, {\it i.e.}~the process responsible for their GZK-cutoff, give rise to a flux of cosmogenic neutrinos~\cite{Stecker:1978ah,Yoshida:pt,Hooper:2004jc} and photons~\cite{Hooper:2010ze}. The accompanying output into secondary electrons and positrons, in particular from Bethe-Heitler pair production, feeds into electromagnetic cascades in the cosmic background radiation and intergalactic magnetic fields~\cite{Protheroe:1992dx}. This leads to the accumulation of $\gamma$-rays at GeV-TeV energies. The observed extra-galactic diffuse $\gamma$-ray flux thus provides a constraint on the total energy injected into such cascades over the Universe's entire history~\cite{Berezinsky:2010xa}.

The methods provided in this paper offer a general tool with which theoretical results may be easily obtained and compared to both these {\it direct} and {\it indirect} UHE CR measurements. As example cases, the application of the general methods developed here to proton propagation provide the opportunity to further develop the method applied in \cite{Taylor:2008jz}. Secondly, an analytic determination of the photon fraction produced through UHE CR nuclei propagation is anticipated to also be obtainable using this treatment. Through the simplicity of our approach and the speed with which it may be implemented, our analytic method is anticipated to be of great benefit as a tool for future UHE CR investigations.

\section*{Acknowledgments}
We thank Jordi Salvad\'o for his help on a numerical evaluation of the cosmic ray spectra via a Runge-Kutta method. M.A.~acknowledges support by the Research Foundation of SUNY at
Stony Brook. 

\appendix

\section{Photo-Disintegration of Nuclei}\label{sec:appI}

\begin{figure}[t]
\centering
\includegraphics[width=0.7\linewidth]{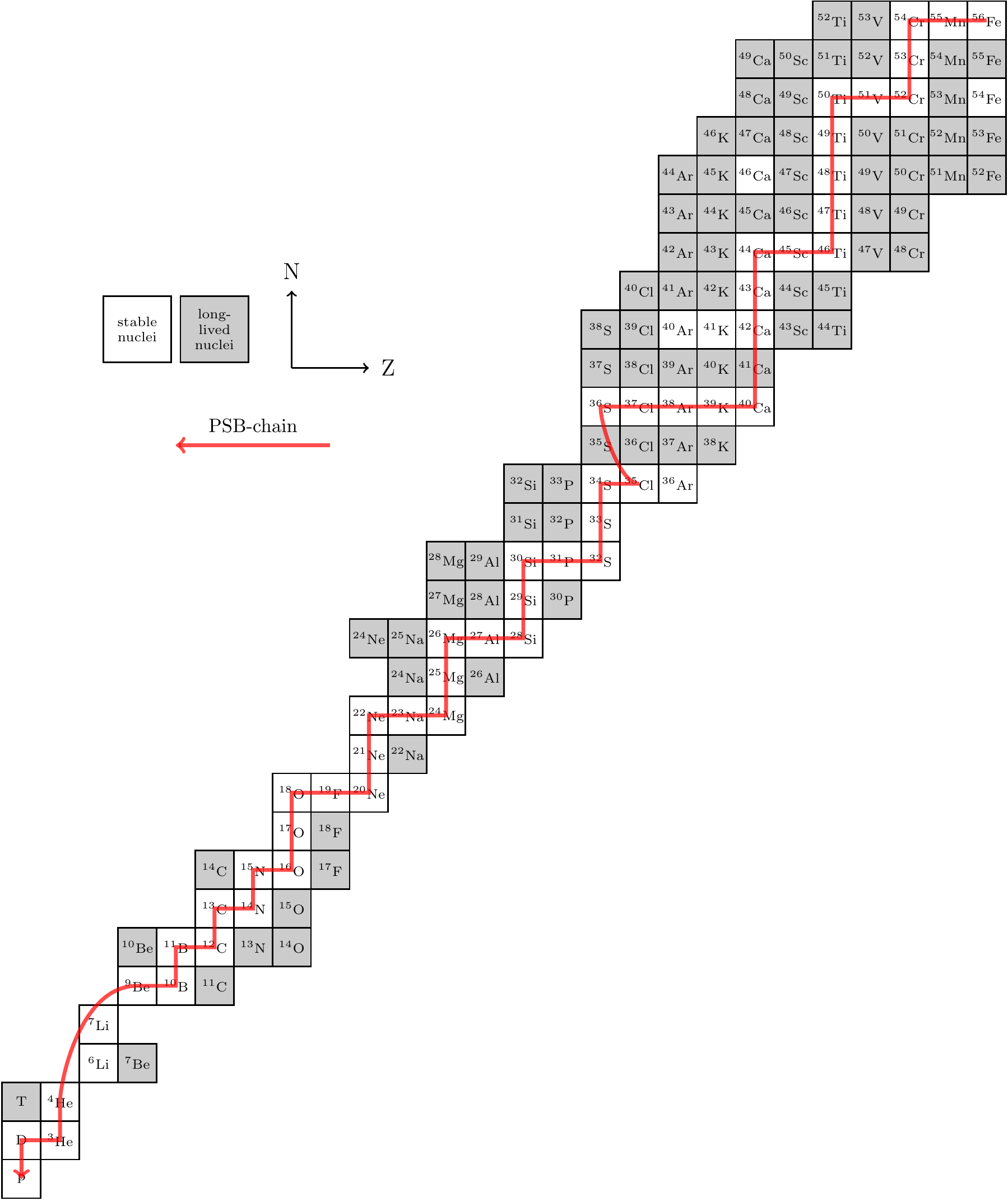}
\caption[]{The {\it Puget-Stecker-Bredekamp}-chain~\cite{Puget:1976nz} along with stable and long-lived ($\tau\gtrsim1$~min) nuclei below ${}^{56}$Fe.}
\label{fig:PSB}
\end{figure}

The most general evolution of primary and secondary nuclei in the CRB includes all possible photo-disintegration transitions between nuclides $(A,Z)$ competing with the decay of unstable nuclides. For simplicity, we follow the work of {\it Puget, Stecker and Bredekamp} (PSB)~\cite{Stecker:1969fw,Puget:1976nz,Stecker:1998ib} and consider only one stable isotope per mass number $A$ in the decay chain of ${}^{56}$Fe as already explained in section~\ref{sec:propagation}. This ``PSB-chain'' is listed in Table~\ref{tab:PSB} and sketched in Fig.~\ref{fig:PSB}.

\begin{table}[t!]\footnotesize
\renewcommand{\arraystretch}{0.5}
\begin{tabular}{l|cc|ccc|cccc|ccc|c}
\hline\hline
\,\,\,nucleus\,\,\,&\,\,\,($\gamma$,n)&($\gamma$,p)\,\,\,&\,\,\,($\gamma$,np)&($\gamma$,2n)&($\gamma$,2p)\,\,\,&\,\,\,($\gamma$,$\alpha$)&($\gamma$,n$\alpha$)&($\gamma$,p$\alpha$)&($\gamma$,2$\alpha$)\,\,\,&\,\,\,($\gamma$,D)&($\gamma$,T)&($\gamma$,${}^3$He)\,\,\,&\,\,\,total\,\,\,\\[2pt]\hline\hline
\,\,\,${}^{56}$Fe&  0.74&  0.12&  0.02&  0.05&  -&  0.02&  -&  -&  -&  -&  -&  -&  0.95\\
\,\,\,${}^{55}$Mn&  0.80&  0.04&  0.01&  0.09&  -&  -&  -&  -&  -&  -&  -&  -&  0.95\\
\,\,\,${}^{54}$Cr&  0.74&  0.01&  0.01&  0.18&  -&  0.01&  -&  -&  -&  -&  -&  -&  0.95\\
\,\,\,${}^{53}$Cr&  0.86&  0.03&  0.02&  0.07&  -&  -&  -&  -&  -&  -&  -&  -&  0.97\\
\,\,\,${}^{52}$Cr&  0.74&  0.11&  0.01&  0.07&  -&  0.02&  -&  -&  -&  -&  -&  -&  0.96\\
\,\,\,${}^{51}$V&  0.79&  0.01&  -&  0.16&  -&  -&  -&  -&  -&  -&  -&  -&  0.96\\
\,\,\,${}^{50}$Ti&  0.79&  0.01&  -&  0.16&  -&  -&  -&  -&  -&  -&  -&  -&  0.96\\
\,\,\,${}^{49}$Ti&  0.84&  0.03&  0.02&  0.08&  -&  -&  -&  -&  -&  -&  -&  -&  0.97\\
\,\,\,${}^{48}$Ti&  0.74&  0.11&  0.02&  0.07&  -&  0.02&  -&  -&  -&  -&  -&  -&  0.95\\
\,\,\,${}^{47}$Ti&  0.81&  0.07&  0.05&  0.02&  -&  0.01&  -&  -&  -&  -&  -&  -&  0.97\\
\,\,\,${}^{46}$Ti&  0.35&  0.46&  0.03&  0.02&  0.03&  0.06&  -&  -&  -&  -&  -&  -&  0.96\\
\,\,\,${}^{45}$Sc&  0.55&  0.27&  0.09&  0.03&  -&  0.02&  -&  -&  -&  -&  -&  -&  0.96\\
\,\,\,${}^{44}$Ca&  0.73&  0.04&  0.01&  0.15&  -&  0.02&  -&  -&  -&  -&  -&  -&  0.95\\
\,\,\,${}^{43}$Ca&  0.74&  0.06&  0.03&  0.08&  -&  0.03&  0.04&  -&  -&  -&  -&  -&  0.97\\
\,\,\,${}^{42}$Ca&  0.37&  0.29&  0.03&  0.07&  0.01&  0.19&  0.01&  -&  -&  -&  -&  -&  0.96\\
\,\,\,${}^{41}$Ca&  0.28&  0.41&  0.13&  -&  0.01&  0.11&  0.01&  -&  -&  -&  -&  -&  0.97\\
\,\,\,${}^{40}$Ca&  0.02&  0.66&  0.02&  -&  0.17&  0.09&  -&  0.01&  -&  -&  -&  -&  0.97\\
\,\,\,${}^{39}$K&  0.08&  0.65&  0.10&  -&  0.01&  0.10&  -&  0.03&  -&  -&  -&  -&  0.98\\
\,\,\,${}^{38}$Ar&  0.46&  0.21&  0.04&  0.05&  0.01&  0.17&  0.01&  -&  -&  -&  -&  -&  0.94\\
\,\,\,${}^{37}$Cl&  0.65&  0.11&  0.04&  0.09&  -&  0.05&  0.01&  -&  -&  -&  -&  -&  0.95\\
\,\,\,${}^{36}$S&  0.68&  0.01&  0.01&  0.23&  -&  0.02&  0.01&  -&  -&  -&  -&  -&  0.96\\
\,\,\,${}^{35}$Cl&  0.12&  0.58&  0.11&  -&  -&  0.13&  -&  0.02&  -&  -&  -&  -&  0.97\\
\,\,\,${}^{34}$S&  0.60&  0.13&  0.03&  0.08&  -&  0.09&  0.01&  -&  -&  -&  -&  -&  0.95\\
\,\,\,${}^{33}$S&  0.44&  0.23&  0.10&  -&  -&  0.13&  0.06&  -&  -&  0.01&  -&  -&  0.97\\
\,\,\,${}^{32}$S&  0.05&  0.63&  0.04&  -&  0.10&  0.14&  -&  0.01&  -&  -&  -&  -&  0.97\\
\,\,\,${}^{31}$P&  0.24&  0.49&  0.13&  -&  -&  0.07&  -&  0.01&  -&  0.01&  -&  -&  0.96\\
\,\,\,${}^{30}$Si&  0.69&  0.04&  0.02&  0.17&  -&  0.03&  0.01&  -&  -&  -&  -&  -&  0.96\\
\,\,\,${}^{29}$Si&  0.65&  0.15&  0.08&  -&  -&  0.06&  0.02&  -&  -&  0.01&  -&  -&  0.97\\
\,\,\,${}^{28}$Si&  0.10&  0.55&  0.04&  -&  0.06&  0.16&  -&  0.01&  -&  -&  -&  -&  0.93\\
\,\,\,${}^{27}$Al&  0.22&  0.44&  0.15&  0.01&  -&  0.10&  -&  0.01&  -&  0.01&  -&  -&  0.94\\
\,\,\,${}^{26}$Mg&  0.68&  0.04&  0.01&  0.17&  -&  0.03&  0.01&  -&  -&  -&  -&  -&  0.95\\
\,\,\,${}^{25}$Mg&  0.64&  0.08&  0.08&  -&  -&  0.10&  0.06&  -&  -&  0.01&  -&  -&  0.97\\
\,\,\,${}^{24}$Mg&  0.08&  0.53&  0.03&  -&  0.03&  0.25&  -&  -&  0.02&  -&  -&  -&  0.96\\
\,\,\,${}^{23}$Na&  0.27&  0.40&  0.10&  0.01&  -&  0.15&  0.01&  -&  0.02&  0.01&  -&  -&  0.96\\
\,\,\,${}^{22}$Ne&  0.65&  0.02&  0.01&  0.17&  -&  0.06&  0.03&  -&  -&  -&  -&  -&  0.95\\
\,\,\,${}^{21}$Ne&  0.49&  0.05&  0.04&  -&  -&  0.21&  0.16&  -&  0.01&  0.01&  -&  -&  0.96\\
\,\,\,${}^{20}$Ne&  0.02&  0.22&  0.02&  -&  -&  0.49&  0.03&  0.06&  0.11&  0.01&  -&  -&  0.96\\
\,\,\,${}^{19}$F&  0.27&  0.12&  0.07&  0.01&  -&  0.35&  0.08&  0.01&  -&  0.01&  0.01&  -&  0.93\\
\,\,\,${}^{18}$O&  0.50&  -&  -&  0.28&  -&  0.09&  0.06&  -&  -&  -&  -&  -&  0.94\\
\,\,\,${}^{17}$O&  0.46&  -&  0.05&  0.01&  -&  0.24&  0.19&  -&  -&  0.01&  -&  -&  0.97\\
\,\,\,${}^{16}$O&  0.09&  0.29&  0.03&  -&  -&  0.36&  -&  0.01&  0.11&  0.02&  -&  -&  0.92\\
\,\,\,${}^{15}$N&  0.38&  0.10&  0.15&  0.02&  -&  0.22&  -&  -&  -&  0.02&  0.02&  -&  0.92\\
\,\,\,${}^{14}$N&  0.15&  0.31&  0.24&  -&  -&  0.10&  0.01&  0.01&  0.01&  0.07&  -&  -&  0.91\\
\,\,\,${}^{13}$C&  0.51&  0.01&  0.03&  0.01&  -&  0.29&  0.13&  -&  -&  -&  -&  -&  0.97\\
\,\,\,${}^{12}$C&  0.11&  0.21&  0.01&  -&  -&  0.57&  0.01&  0.02&  -&  0.01&  -&  -&  0.94\\
\,\,\,${}^{11}$B&  0.21&  0.05&  0.04&  0.01&  -&  0.32&  0.10&  -&  -&  0.05&  0.11&  -&  0.89\\
\,\,\,${}^{10}$B&  0.14&  0.21&  0.03&  -&  -&  0.38&  -&  0.01&  -&  0.17&  0.01&  0.01&  0.96\\[2pt]
\hline\hline
\end{tabular}
\caption[]{The nuclei of the {\it Puget-Stecker-Bredekamp}-chain~\cite{Puget:1976nz} and the relative contribution of inclusive channels to the total photo-disintegration cross section in the CMB calculated by {\tt TALYS}~\cite{Goriely:2008zu}. We assume and $E^{-2}$ spectrum of the nuclei and integrate over nucleon energies $10^{17}{\rm eV}<E/A<10^{21}{\rm eV}$. Channels with contribution less than 1\% are omitted in the table.}
\label{tab:PSB}
\end{table}

Table~\ref{tab:PSB} shows the relative contribution of inclusive channels to the total photo-disintegration rate calculated for the nuclei of the PSB-chain.  We use the reaction code {\tt TALYS}~\cite{Goriely:2008zu} to evaluate the cross sections for nuclei with $10\leq A\leq56$ and assume an $E^{-2}$ power-law flux of CR nuclei. At CR energies $E<10^{12}$~GeV and large mass numbers $A\gtrsim 20$ photo-disintegration in the CRB can be well approximated by one-nucleon and two-nucleon losses between elements of the PSB-chain via exclusive processes $(\gamma,p)$, $(\gamma,n)$, $(\gamma,2p)$, $(\gamma,2n)$ and $(\gamma,pn)$. For the cross sections of light nuclei with mass numbers $A=2,3,4$ and $9$ we use the parameterization of Ref.~\cite{RachenTHESIS}.

At lower mass numbers, $A\lesssim 20$, additional channels involving $\alpha$ particle emission can become as significant as the sume of one-nucleon and two-nucleon losses. Table~\ref{tab:PSB} also shows the relative importance of the exclusive channels $(\gamma,\alpha)$, $(\gamma,n\alpha)$, $(\gamma,p\alpha)$ and $(\gamma,2\alpha)$ to the total photo-disintegration budget. Resonant photo-nuclear interactions play only a minor role in the propagation of the nuclei for the energies of interest. We follow the approach outlined in Ref.~\cite{RachenTHESIS} and approximate the total interaction by the isospin averaged $N\gamma$ rate as $\Gamma_{A\gamma}(z,E) \simeq A \Gamma_{N\gamma}(z,E/A)$. We also assume that the participating nucleon is removed from the nucleus and regard this as a contribution to one-nucleon losses.

The angle-averaged interaction rate appearing in Eq.~(\ref{eq:diff0}) is then defined as
\begin{equation}
\Gamma_{A\to B}(z,E) =
\frac{1}{2}\int\limits_{-1}^1\mathrm{d}\cos\theta\int\mathrm{d}
\epsilon\,(1-\beta
\cos\theta) n_\gamma(z,\epsilon)\sigma_{A\to B}(\epsilon')\,,
\end{equation}
where $n_\gamma(z,\epsilon)$ is the energy distribution of isotropic background
photons at red-shift $z$ and $\epsilon'=\epsilon\gamma(1-\beta
\cos\theta)$ the photon's energy in the rest frame of the nucleus. For our calculation we use the cosmic microwave background and the infra-red/optical background form Ref.~\cite{Franceschini:2008tp}. 
To a good approximation the decay products of the photo-disintegration interaction inherit the large boost-factor of the initial nucleus and hence in the process $A\to B + (A-B)$ the nucleus with mass number $B$ has an energy $E' = (B/A)E$. We can hence approximate the differential cross section as 
\begin{equation}\label{eq:gammaAB}
\gamma_{A\to B}(E,E') \simeq \Gamma_{A\to B}(E)\delta((B/A)E-E')
\end{equation} 
in the following. This has the correct normalization since $\Gamma_{A\to B}(E) \equiv\int {\rm d}E'\gamma_{A\to B}(E,E')$.

In general, the interaction rates $\Gamma_{A\to B}(z,E)$ scale with red-shift according to the red-shift evolution of the radiation background. In the case of the CMB with adiabatically scaling, $n_\gamma(z,\epsilon) = (1+z)^2\,n_\gamma(0,\epsilon/(1+z))$, we can derive the simple relation
\begin{equation}\label{scaling1}
\Gamma_{A\to B}(z,E_i) = (1+z)^3\,\Gamma_{A\to B}(0,(1+z)E)\,.
\end{equation}
For the case of the infra-red/optical background~\cite{Franceschini:2008tp} we assume a red-shift scaling following the star formation rate as described in Ref.~\cite{Ahlers:2009rf}. However, since the cascades of UHE CR nuclei develop locally, the red-shift dependence of the interaction rates is only of minor importance.

\section{Proof of Equation (\ref{eq:fullsol})}\label{sec:appII}

We will proof Eq.~(\ref{eq:fullsol}) by induction. First note, that we can rewrite Eq.(\ref{eq:fullsol}) as
\begin{equation}\label{eq:deqproof}
\dot N_i = -\Gamma^{\rm tot}_i N_i + \sum_{j=i+1}^n\Gamma_{j\to i}N_j + Q_i\,,
\end{equation}
with $i=1,\ldots,n$ with $\Gamma_{j\to i}=0$ for $i\leq j$ and $\Gamma^{\rm tot}_i\neq\Gamma^{\rm tot}_j$ for $i\neq j$. In the following we will refer to the indices $i$ as {\it knots} and the pairs $(i,j)$ with $\Gamma_{i\to j}\neq0$ as {\it links}. A {\it chain} of length $n_c$ is defined as an ascending sequence of $n_c$ knots, $c_1<c_2<\ldots<c_{n_c}$, that are mutually connected by links.

We want to show that the most general solution of Eq.~(\ref{eq:deqproof}) is of the form
\begin{equation}\label{eq:fullsol2}
N_{i}(t) = \sum_{j\leq i}\sum_{\mathbf{c}}\left(\prod_{l=1}^{n_c-1}\Gamma_{c_{l}\to c_{l+1}}\right)\sum_{k=1}^{n_c}\left[N_{j}(0)e^{-t\Gamma^{\rm tot}_{c_k}}+\int_0^t{\rm d}t'Q_j(t')e^{-(t-t')\Gamma^{\rm tot}_{c_k}}\right]\prod_{p=1(\neq k)}^{n_c}\frac{1}{\Gamma^{\rm tot}_{c_p}-\Gamma^{\rm tot}_{c_k}}
\end{equation}
where the sum is over all possible chains $\mathbf{c}$ with $c_{1} = j$ and $c_{n_c} = i$.

\noindent\emph{Induction start: $n=1$.} This case has the solution 
\[
N_1(t) = N_1(0)e^{-t\Gamma^{\rm tot}_1}+\int_0^t{\rm d}t'Q_1(t')e^{-(t-t')\Gamma^{\rm tot}_1}\,.
\]
This is of the form (\ref{eq:fullsol2}), since the only chain is the trivial one of length $n_c=1$ with $c_1 = 1$.

\noindent\emph{Induction step: $n\to n+1$.} The differential equations of $N_i$ with $1\leq i\leq n$ are of the form (\ref{eq:deqproof}) and we can hence use the solution (\ref{eq:fullsol2}).
The differential equation for $N_{n+1}$ is
\begin{equation}
\dot N_{n+1} = -\Gamma^{\rm tot}_{n+1} N_{n+1} + \sum_{m=1}^n\Gamma_{m\to n+1}N_m + Q_{n+1}\,,
\end{equation}
We can write the general solution of this differential equation as:
\begin{equation}
N_{n+1}(t) = \left[N_{n+1}(0)e^{-t\Gamma^{\rm tot}_{n+1}}+\int_0^t{\rm d}t'Q_{n+1}(t')e^{-(t-t')\Gamma^{\rm tot}_{n+1}}\right] + \int_0^t{\rm d}t'e^{-(t-t')\Gamma^{\rm tot}_{n+1}}\sum_{m=1}^n\Gamma_{m\to n+1}N_m(t')\,.
\end{equation}
The first term of the previous equation corresponds to the first term ($i=j=n+1$) in the sum of Eq.(\ref{eq:fullsol2}). 
Inserting the solutions (\ref{eq:fullsol2}) in the integrand yields after integration by parts:
\begin{align}\label{eq:aux2}
N_{n+1}(t) &= \left[N_{n+1}(0)e^{-t\Gamma^{\rm tot}_{n+1}}+\int_0^t{\rm d}t'Q_{n+1}(t')e^{-(t-t')\Gamma^{\rm tot}_{n+1}}\right]\\&+\sum_{m=1}^n\sum_{j=1}^m\sum_{\mathbf{c}}\left(\prod_{l=1}^{n_c-1}\Gamma_{c_{l}\to c_{l+1}}\right)\Gamma_{m\to n+1}\sum_{k=1}^{n_c}\left(\prod_{p=1(\neq k)}^{n_c}\frac{1}{\Gamma^{\rm tot}_{c_p}-\Gamma^{\rm tot}_{c_k}}\right)\frac{1}{\Gamma^{\rm tot}_{n+1}-\Gamma^{\rm tot}_{c_k}}\nonumber\\&\qquad\qquad\qquad\times\left(\left[N_{j}(0)e^{-t\Gamma^{\rm tot}_{c_k}}+\int_0^t{\rm d}t'Q_j(t')e^{-(t-t')\Gamma^{\rm tot}_{c_k}}\right]-\left[N_{j}(0)e^{-t\Gamma^{\rm tot}_{{n+1}}}+\int_0^t{\rm d}t'Q_j(t')e^{-(t-t')\Gamma^{\rm tot}_{n+1}}\right]\right)\,.\nonumber
\end{align}
The chains $\mathbf{c}$ in the previous sums have end-points $c_1=j$ and $c_{n_c}=m$. Now, every chain $\mathbf{c}$ in the system with $n$ knots and endpoint $c_{n_c}=m$ corresponds {\it unambiguously} to a chain $\mathbf{c}'$ in the system with $n+1$ knots with $c'_i=c_i$ for $i\leq n_{c'}-1$ and $c'_{n_{c'}}=n+1$.
Hence, the double-sum in Eq.(\ref{eq:aux2}) over end-points $m<n$ and chains $\mathbf{c}$ can be expressed as a single sum over chains $\mathbf{c}'$ with $c'_1=j$ and $c'_{n_{c'}}=n+1$. We arrive at the form:
\begin{align}\label{eq:aux3}
N_{n+1}(t) &= \left[N_{n+1}(0)e^{-t\Gamma^{\rm tot}_{n+1}}+\int_0^t{\rm d}t'Q_{n+1}(t')e^{-(t-t')\Gamma^{\rm tot}_{n+1}}\right]\\&+\sum_{j=1}^n\sum_{\mathbf{c'}}\left(\prod_{l=1}^{n_{c'}-1}\Gamma_{c'_{l}\to c'_{l+1}}\right)\sum_{k=1}^{n_c'-1}\left[N_{j}(0)e^{-t\Gamma^{\rm tot}_{c'_k}}+\int_0^t{\rm d}t'Q_j(t')e^{-(t-t')\Gamma^{\rm tot}_{c'_k}}\right]\prod_{p=1(\neq k)}^{n_{c'}}\frac{1}{\Gamma^{\rm tot}_{c'_p}-\Gamma^{\rm tot}_{c'_k}}\nonumber\\
&-\sum_{j=1}^n\sum_{\mathbf{c}'}\left(\prod_{l=1}^{n_{c'}-1}\Gamma_{c'_{l}\to c'_{l+1}}\right)\left[N_{j}(0)e^{-t\Gamma^{\rm tot}_{{n+1}}}+\int_0^t{\rm d}t'Q_j(t')e^{-(t-t')\Gamma^{\rm tot}_{n+1}}\right]\sum_{k=1}^{n_{c'}-1}\prod_{p=1(\neq k)}^{n_{c'}}\frac{1}{\Gamma^{\rm tot}_{c'_p}-\Gamma^{\rm tot}_{c'_k}}\,.\nonumber
\end{align}
As a final step we use the identity:\footnote{See the appendix of Ref.~\cite{Hooper:2008pm} for a simple derivation of this expression.}
\begin{equation}
\sum_{k=1}^{n_{c'}-1}\prod_{p=1(\neq k)}^{n_{c'}}\frac{1}{\Gamma^{\rm tot}_{c'_p}-\Gamma^{\rm tot}_{c'_k}}=-\prod_{p=1}^{n_{c'}-1}\frac{1}{\Gamma^{\rm tot}_{c'_p}-\Gamma^{\rm tot}_{n+1}}\,,
\end{equation} 
to combine the last two terms in Eq.~(\ref{eq:aux3}) and arrive at the form (\ref{eq:fullsol2}).\hfill$\Box$


\begin{thebibliography}{99}

\bibitem{Nagano:2000ve}
  M.~Nagano and A.~A.~Watson,
  Rev.\ Mod.\ Phys.\  {\bf 72}, 689 (2000).

\bibitem{Abraham:2010mj}
  J.~Abraham {\it et al.}  [Pierre Auger Collaboration],
  Phys.\ Lett.\  B {\bf 685}, 239 (2010).

\bibitem{Abraham:2010yv}
  J.~Abraham {\it et al.}  [Pierre Auger Collaboration],
  Phys.\ Rev.\ Lett.\  {\bf 104}, 091101 (2010).
  
\bibitem{Abbasi:2007sv}
  R.~Abbasi {\it et al.}  [HiRes Collaboration],
  Phys.\ Rev.\ Lett.\  {\bf 100}, 101101 (2008).
  
\bibitem{Abbasi:2009nf}
  R.~U.~Abbasi {\it et al.}  [HiRes Collaboration],
  Phys.\ Rev.\ Lett.\  {\bf 104}, 161101 (2010).
  
\bibitem{Linsley:1963bk}
  J.~Linsley, {\it Proceedings of ICRC 1963, Jaipur, India}, pp.~77-99

\bibitem{Hill:1983mk}
  C.~T.~Hill and D.~N.~Schramm,
  Phys.\ Rev.\  D {\bf 31}, 564 (1985).

\bibitem{Berezinsky:2002nc}
  V.~Berezinsky, A.~Z.~Gazizov and S.~I.~Grigorieva,
  Phys.\ Rev.\  D {\bf 74}, 043005 (2006).
  
\bibitem{Fodor:2003ph} 
Z.~Fodor, S.~D.~Katz, A.~Ringwald and H.~Tu,
JCAP {\bf 0311}, 015 (2003).

\bibitem{Greisen:1966jv}
  K.~Greisen,
  Phys.\ Rev.\ Lett.\  {\bf 16}, 748 (1966).

\bibitem{Zatsepin:1966jv}
  G.~T.~Zatsepin and V.~A.~Kuzmin,
  JETP Lett.\  {\bf 4}, 78 (1966)
  [Pisma Zh.\ Eksp.\ Teor.\ Fiz.\  {\bf 4}, 114 (1966)].

\bibitem{Abraham:2008ru}
  J.~Abraham {\it et al.}  [Pierre Auger Collaboration],
  Phys.\ Rev.\ Lett.\  {\bf 101}, 061101 (2008).
 
\bibitem{Amsler:2008zzb}
  C.~Amsler {\it et al.}  [Particle Data Group],
  Phys.\ Lett.\  B {\bf 667}, 1 (2008).
  
\bibitem{Hooper:2008pm}
  D.~Hooper, S.~Sarkar and A.~M.~Taylor,
  Phys.\ Rev.\  D {\bf 77}, 103007 (2008).

\bibitem{Aloisio:2008pp}
  R.~Aloisio, V.~Berezinsky and S.~Grigorieva,
 arXiv:0802.4452 [astro-ph];
  arXiv:1006.2484 [astro-ph].
  
\bibitem{Ahlers:2009rf}
  M.~Ahlers, L.~A.~Anchordoqui and S.~Sarkar,
  Phys.\ Rev.\  D {\bf 79}, 083009 (2009).

\bibitem{Stecker:1969fw}
  F.~W.~Stecker,
  Phys.\ Rev.\  {\bf 180}, 1264 (1969).

\bibitem{Puget:1976nz}
  J.~L.~Puget, F.~W.~Stecker and J.~H.~Bredekamp,
  Astrophys.\ J.\  {\bf 205}, 638 (1976).

\bibitem{Stecker:1998ib}
  F.~W.~Stecker and M.~H.~Salamon,
  Astrophys.\ J.\  {\bf 512}, 521 (1999).
  
\bibitem{Goriely:2008zu}
  S.~Goriely, S.~Hilaire and A.~J.~Koning,
  Astron.\ Astrophys.\  {\bf 487}, 767 (2008), 
  \url{http://www.talys.eu/}
  
\bibitem{Blumenthal:1970nn}
  G.~R.~Blumenthal,
  Phys.\ Rev.\  D {\bf 1}, 1596 (1970).

\bibitem{Kronberg:1993vk}
  P.~P.~Kronberg,
  Rept.\ Prog.\ Phys.\  {\bf 57}, 325 (1994).

\bibitem{Neronov:1900zz}
  A.~Neronov and I.~Vovk,
  Science {\bf 328}, 73 (2010).

\bibitem{Dolag:2004kp}
  K.~Dolag, D.~Grasso, V.~Springel and I.~Tkachev,
  JCAP {\bf 0501}, 009 (2005).
  
\bibitem{Aloisio:2004jda}
  R.~Aloisio and V.~Berezinsky,
  Astrophys.\ J.\  {\bf 612}, 900 (2004).

\bibitem{Aloisio:2004fz}
  R.~Aloisio and V.~S.~Berezinsky,
  Astrophys.\ J.\  {\bf 625}, 249 (2005).

\bibitem{GMP}
{\it {GNU} Multiple Precision Arithmetic Library 5.0.1}, \url{http://gmplib.org/}

\bibitem{MPFR}
{\it {GNU} Multiple Precision Floating-Point Reliable Library 3.0.0}, \url{http://www.mpfr.org/}

\bibitem{GSL}
{\it {GNU} Scientific Library 1.14}, \url{http://www.gnu.org/software/gsl/}

\bibitem{Mucke:1999yb}
  A.~M\"ucke, R.~Engel, J.~P.~Rachen, R.~J.~Protheroe and T.~Stanev,
  Comput.\ Phys.\ Commun.\  {\bf 124}, 290 (2000). 

\bibitem{Tokuno:2010zz}
  H.~Tokuno {\it et al.},
  AIP Conf.\ Proc.\  {\bf 1238} (2010) 365.
    
\bibitem{Ahlers:2005sn}
  M.~Ahlers, L.~A.~Anchordoqui, H.~Goldberg, F.~Halzen, A.~Ringwald and T.~J.~Weiler,
  Phys.\ Rev.\  D {\bf 72}, 023001 (2005).
  
\bibitem{Stecker:1978ah}
  F.~W.~Stecker,
  Astrophys.\ J.\  {\bf 228}, 919 (1979).

\bibitem{Yoshida:pt}
S.~Yoshida and M.~Teshima,
Prog.\ Theor.\ Phys.\  {\bf 89}, 833 (1993);
R.~J.~Protheroe and P.~A.~Johnson,
Astropart.\ Phys.\  {\bf 4}, 253 (1996); 
R.~Engel, D.~Seckel and T.~Stanev,
Phys.\ Rev.\ D {\bf 64}, 093010 (2001).

\bibitem{Hooper:2004jc}
D.~Hooper, A.~Taylor and S.~Sarkar,
Astropart.\ Phys.\  {\bf 23}, 11 (2005);
M.~Ave, N.~Busca, A.~V.~Olinto, A.~A.~Watson and T.~Yamamoto,
Astropart.\ Phys.\  {\bf 23}, 19 (2005);
 D.~Allard {\it et al.},
  JCAP {\bf 0609}, 005 (2006);
  L.~A.~Anchordoqui, H.~Goldberg, D.~Hooper, S.~Sarkar and A.~M.~Taylor,
  Phys.\ Rev.\  D {\bf 76}, 123008 (2007);
  K.~Kotera, D.~Allard and A.~V.~Olinto,
  arXiv:1009.1382 [astro-ph.HE].

\bibitem{Hooper:2010ze}
  D.~Hooper, A.~M.~Taylor and S.~Sarkar,
  arXiv:1007.1306 [astro-ph.HE].

\bibitem{Protheroe:1992dx}
  R.~J.~Protheroe and T.~Stanev,
  Mon. Not. R. Astron. Soc. {\bf 264}, 191 (1993);
  S.~Lee,
  Phys.\ Rev.\  D {\bf 58}, 043004 (1998).

\bibitem{Berezinsky:2010xa}
 For recent evalutations see, V.~Berezinsky, A.~Gazizov, M.~Kachelriess and S.~Ostapchenko,
  arXiv:1003.1496 [astro-ph.HE];
  M.~Ahlers, L.~A.~Anchordoqui, M.~C.~Gonzalez-Garcia, F.~Halzen and S.~Sarkar,
  Astropart.\ Phys.\  {\bf 34}, 106 (2010).

\bibitem{Taylor:2008jz}
  A.~M.~Taylor and F.~A.~Aharonian,
  Phys.\ Rev.\  D {\bf 79}, 083010 (2009). 

\bibitem{RachenTHESIS}
 J.~P.~Rachen, {\it Interaction processes and statistical properties of the propagation of cosmic-rays in photon 
backgrounds}, PhD thesis of the Bonn University, 1996. 
  
\bibitem{Franceschini:2008tp}
 A.~Franceschini, G.~Rodighiero and M.~Vaccari,
 Astron.\ Astrophys.\  {\bf 487}, 837 (2008).
 
\end{thebibliography}
\end{document}